\documentclass[12pt,preprint]{aastex}
\usepackage{subfigure}

\IfFileExists{srcltx.sty}{\usepackage[active]{srcltx}}

\shorttitle{SECONDARY GAMMA RAYS AND NEUTRINOS FROM BLAZARS}
\shortauthors{BEACOM et al.}

\begin{document}

\title{Role of line-of-sight cosmic ray interactions in forming the spectra of distant blazars in TeV gamma rays 
and high-energy neutrinos}

\author{Warren~Essey\altaffilmark{1},
 Oleg~Kalashev\altaffilmark{2}, Alexander Kusenko \altaffilmark{1,3}, John F. Beacom\altaffilmark{4,5,6}}

\altaffiltext{1}{Department of Physics and Astronomy, University of
California, Los Angeles, CA 90095-1547, USA}
\altaffiltext{2}{Institute for Nuclear Research, 60th October Anniversary Prospect 7a, Moscow 117312 Russia}
\altaffiltext{3}{Institute for the Physics and Mathematics of the Universe,
University of Tokyo, Kashiwa, Chiba 277-8568, Japan}
\altaffiltext{4}{Center for Cosmology and Astro-Particle Physics, Ohio State University, Columbus, Ohio 43210, USA}
\altaffiltext{5}{Department of Physics, Ohio State University, Columbus, Ohio 43210, USA}
\altaffiltext{6}{Department of Astronomy, Ohio State University, Columbus, Ohio 43210, USA}

\begin{abstract}
Active galactic nuclei (AGN) can produce both gamma rays and cosmic rays. The observed high-energy gamma-ray signals from distant blazars may be dominated by secondary gamma rays produced along the line of sight by the interactions of cosmic-ray protons with background photons. This explains the surprisingly low attenuation observed for distant blazars, because the production of secondary gamma rays occurs, on average, much closer to Earth than the distance to the source. Thus the observed spectrum in the TeV range does not depend on the intrinsic gamma-ray spectrum, while it depends on the output of the source in cosmic rays.  We apply this hypothesis to a number of sources and, in every case, we obtain an excellent fit, strengthening the interpretation of the 
observed spectra as being due to secondary gamma rays.  We explore the ramifications of this interpretation for limits on the extragalactic background light and for the production of cosmic rays in AGN.  We also make predictions for the neutrino signals, which can help probe the acceleration of cosmic rays in AGN.
\end{abstract}

\keywords{gamma rays, cosmic rays, active galaxies}


 \section{Introduction}

Active galactic nuclei (AGN) are believed to produce both gamma rays and cosmic rays. Gamma-ray photons with energies in the TeV range have been observed from a number of distant blazars~\citep{Aharonian:2005gh,2006ApJ...639..761A,2007A&A...473L..25A,2009ApJ...693L.104A,2008AIPC.1085..644C}.  It is believed that blazars also accelerate high-energy cosmic rays, but 
it is more difficult, if at all possible, to associate the arrival directions of cosmic rays with extragalactic sources because of the effects of the Milky Way magnetic fields~\citep{Harari:2000he,Harari:2000az,Cronin:2007zz,Abraham:2007si,Golup:2009cv,Murase:2010va,Calvez:2010uh}, which are known to have magnitudes as large as several
microgauss.   Gamma-ray spectra of distant sources have been used as a probe of universal photon   backgrounds~\citep{Stecker:1992wi,1994Natur.369..294D,1998ApJ...493..547S,1998ApJ...494L.159S,Aharonian:2005gh,Mazin:2007pn,2009ApJ...698.1761F,2010A&A...522A..12Y,2010ApJ...723.1082A}. 
However, it was recently pointed out that the highest-energy gamma rays from the more distant objects may be secondary gamma rays produced relatively close to Earth in proton-photon interactions, and not primary gamma rays emitted at the source~\citep{Essey:2009zg,Essey:2009ju}. 
Although the energy attenuation length of protons below the Greisen-Zatsepin-Kuzmin (GZK) cutoff~\citep{Greisen:1966jv,Zatsepin:1966jv} is much larger than the horizon size, the protons do rarely interact with the background photons and produce secondary gamma rays at a rate that results in observable flux and shows an excellent agreement with the data~\citep{Essey:2009zg,Essey:2009ju}.  The contribution of cosmic rays to the observed spectra of distant blazars should not be neglected in analyses of gamma-ray opacity of the universe and in deriving constraints on the photon backgrounds. 

The identification of observed gamma rays with secondary showers along the line of sight reconciles the observed TeV spectra with theoretical models.  The primary gamma rays should exhibit clear signatures of absorption due to their interactions with extragalactic background light (EBL), such as a sharp cutoff at energies around 1~TeV that was predicted prior to observations~\citep{Stecker:1992wi}.  However, the observed spectra do not show such features~\citep{Aharonian:2005gh,2009ApJ...693L.104A,2008AIPC.1085..644C}.  This can be explained by the lower levels of EBL~\citep{Aharonian:2005gh,Mazin:2007pn,2009ApJ...698.1761F}, combined with much harder intrinsic spectra of distant blazars~\citep{2007ApJ...667L..29S} than those predicted by earlier models~\citep{1981MNRAS.196..135P,1987ApJ...315..425K,1988MNRAS.235..997H,1998PhRvL..80.3911B,2001RPPh...64..429M}.  Alternatively, the lack of absorption features can be explained by the production of secondary gamma rays, which replace the primary gamma-rays lost to their interactions with EBL~\citep{Essey:2009zg,Essey:2009ju}.  We will present additional evidence that the latter approach provides a consistent explanation of all the present data.  

Secondary gamma rays produce point images as long as the magnetic deflections of protons are sufficiently small. The magnetic fields in the source host galaxy introduce only a negligible image broadening, at most, of the order of the galaxy size divided by the distance to the source.  In intergalactic space, the magnetic fields are much weaker than inside a galaxy, and can have the field strenghths as low as a femtogauss~\citep{Ando:2010rb,Kandus:2010nw}. Therefore, the proton deflections are small, and the images of sources in secondary photons appear pointlike.  Indeed, the deflections of protons emitted by a distant blazar on their passage in a random magnetic field below 10~fG are smaller than the angular resolution of atmospheric Cherenkov telescopes (ACT): 
\begin{equation}
 \Delta \theta \sim 0.1^{\circ}
\left( \frac{B}{10^{-14}\rm G}\right)
\left( \frac{4\times 10^{7} {\rm GeV}}{E} \right)
 \left( \frac{D}{1\, \rm Gpc}\right)^{1/2}
\left( \frac{l_{\rm c} }{1\, \rm Mpc}\right)^{1/2}.
\label{deflections}
\end{equation}
Here $D$ is the distance to the source $D$ and  $l_{\rm c}$ is the average correlation length. 

It is easy to understand qualitatively why, for a sufficiently distant source, secondary gamma-ray flux should dominate over primary gamma-ray flux.  
The flux of the primary gamma rays is attenuated by their interactions with EBL and, for a source at distance $d$, the flux of unattenuated high-energy gamma rays is 
\begin{equation}
F_{\rm primary, \gamma}(d) \propto \frac{1}{d^2} \exp\{-d/\lambda_\gamma\} , 
\label{primary}
\end{equation}
where $\lambda_\gamma$ is the attenuation length due to the interactions with EBL. 
Cosmic rays emitted from the same source with energies below the GZK cutoff~\citep{Greisen:1966jv,Zatsepin:1966jv} of about $3\times 10^{10}$~GeV can cross cosmological distances without a significant energy loss.  Their flux at distance $d$ is $F_{\rm protons} \propto 1/d^2$.  Although the photon background is optically thin for cosmic rays, the protons do rarely interact with the cosmic background photons and produce gamma rays.  

Let us consider an isotropic source of protons.  Since the proton flux is not attenuated, the same number of protons pass through every spherical shell at any radius $r$.  Let protons produce gamma rays at the rate $p$ per unit length of path, and the number of gamma rays passing through a shell at distance $r$ is $\Phi_\gamma (r)$.  Some gamma rays are lost to pair production characterized by attenuation length $\lambda_\gamma$.  The change in the number of photons $\Phi_\gamma$ that occurs between $r$ and $r+dr$ is $d\Phi_\gamma = p\, dr -(1/\lambda_\gamma) \Phi_\gamma \, dr$.  The solution of this equation with a boundary condition $\Phi_\gamma(0)=0$, appropriate for secondary gamma rays, is $\Phi_\gamma(r)=p\lambda_\gamma \left [1-\exp(-r/\lambda_\gamma)  \right ]$. For the flux per unit area $F_{{\rm secondary,} \, \gamma}(d) =\Phi_\gamma (d) /(4\pi d^2)$, this solution gives 
\begin{equation}
 F_{{\rm secondary}, \gamma}(d) = \frac{p\lambda_\gamma}{4\pi d^2}
\left [1-e^{-d/\lambda_\gamma}  \right ]
\propto 
\left \{ 
\begin{array}{ll}
1/d, & {\rm for} \ d \ll \lambda_\gamma, \\ 
1/d^2, & {\rm for} \ d\gg \lambda_\gamma .
\end{array} \right.
\label{secondary_photons}
\end{equation}
This derivation extends to beamed sources, as long as the effects of the beam broadening are small, which is the case for IGMF below 10~fG, in agreement with observational data~\citep{Kandus:2010nw}. 

It is clear from Eqns.~(\ref{primary}-\ref{secondary_photons}) that, for a sufficiently high proton flux, secondary gamma rays should dominate the spectra of very distant sources 
above $E\sim $TeV, because their flux suppression is less  severe at large distances.  

Secondary neutrinos also show a different scaling with distance.  This can be used as a tool to distinguish between primary neutrinos from AGN~\citep{Stecker:1991vm} and secondary neutrinos produced along the line of sight~\citep{Essey:2009ju}.  For neutrinos, there is no absorption, and the flux of secondary neutrinos scales as 
\begin{equation}
F_{{\rm secondary}, \nu} (d) \propto \left( F_{\rm protons} \times d \right) \propto \frac{1}{d}. 
\label{secondary_neutrinos}
\end{equation}
The 1/d scaling applies as long as the intergalactic magnetic fields (IGMF) are sufficiently small to allow the protons to remain within the angular resolution of the detector. This unique scaling law can be exploited by future experiments: a larger number of sources should be within the field of view than one would predict based on the primary gamma-ray flux.

The secondary gamma rays should arrive at random times and show no temporal variability, and this is consistent with the data.  While variability has been observed for {\em nearby} TeV blazars at TeV energies~\citep{Krennrich:2002as,Blazejowski:2005ih,2006Sci...314.1424A} and for distant TeV blazars at energies above 200~GeV~\citep{2009ApJ...693L.104A,2010ApJ...709L.163A}, 
no variability has been observed for {\em distant} TeV blazars at {\em TeV or higher energies}.   The flux of gamma rays with $E>200$~GeV is dominated by the photons with energies $E\approx 200$~GeV, not TeV.  We expect to see no variability above TeV for blazars with $z>0.1$ considered here.  

The current data are consistent with the interpretation that secondary gamma rays dominate observed signals from distant sources at the highest energies~\citep{Essey:2009zg,Essey:2009ju}.

Of course, the relations in Eqns.~(\ref{primary}-\ref{secondary_neutrinos}) are crude approximations, and one must take into account a complex chain of interactions that ensues when the first interaction triggers an electromagnetic cascade.  To this end, we will present the results of numerical Monte Carlo simulations that account for all the relevant interactions and we will compare theoretical predictions to the data. The overall normalization is uncertain, but the shape of the spectrum is fixed.  Hence, our model predictions should be viewed as one-parameter fits to the data.  In all the cases we find an excellent agreement between the shape of the spectrum predicted by the model and the data.  We will discuss the implications of our results for studies of extragalactic background light and other areas where it is important to distinguish between primary and secondary gamma rays.  In what follows we will assume that IGMFs are smaller than 10~fG; this assumption is in agreement with all the present observational data~\citep{Kandus:2010nw}, and it allows us to explore the role of line-of-sight cosmic-ray interactions in the case where they dominate, while the highest-energy intrinsic gamma rays originating at the source are filtered out.  The effects of larger magnetic fields were discussed by~\cite{Essey:2010nd}.

\section{Gamma rays from distant blazars}

To obtain the spectra that can be compared with the data, we employed a detailed Monte-Carlo simulation, tracking individual cosmic ray protons and all secondary particles from the source through the intergalactic medium. The intergalactic medium was modeled by a background photon field, consisting of the CMB and EBL, and IGMF. 

We have considered a wide range of EBL models, including the model with the lowest level of background light based on the lower limits obtained from galaxy counts~\citep{2001ApJ...562..179X} and the highest level based on~\cite{2006ApJ...648..774S}. The IGMF was modeled by 3D cubes of the size of a characteristic correlation length, with the direction randomly chosen for each cube.

The dominant interactions for the protons are proton pair production (PPP) and pion photoproduction: 
\begin{eqnarray}
p+\gamma_b & \rightarrow & p + e^+ + e^- \label{PPP} \\
p+\gamma_b & \rightarrow & n+\pi^+  \label{photopion:+} \\
p+\gamma_b & \rightarrow & p+\pi^0 \label{photopion:0} 
\end{eqnarray}
where $\gamma_b$ is a background photon. Neutrons and pions subsequently decay and produce neutrinos, photons, electrons and positrons. 
Individual proton interactions and EM cascades were modeled using a standard Monte Carlo approach where energies and directions were sampled from distribution functions constructed from the appropriate cross sections~\citep{1994APh.....2..375S,1970PhRvD...1.1596B,1986MNRAS.221..769P}. The outgoing distribution functions for pion photoproduction were generated using the SOPHIA package~\citep{2000CoPhC.124..290M}.

For each iteration, particles were propagated a distance far less than the average correlation length of the magnetic field, to ensure the accuracy in calculated deflections. Two cuts were applied to the particles arriving at the $z=0$ surface to decide whether or not to include them in the observed spectrum. First, the particle must point back to a point in the sky that is within an angular distance defined by the point spread function (PSF) of the observing instrument. For energies below 100~GeV the Fermi PSF~\citep{2009arXiv0907.0626R} was used, and for energies above 100~GeV a PSF for a typical atmospheric  Cherenkov telescope such as HESS, MAGIC, or VERITAS~\citep{2008AIPC.1085..657H} was used. Second, the particle must arrive within a cone that is characterized by the jet opening angle for the source.

\begin{figure}
\begin{center}

\begin{tabular}{cc}
\subfigure[\label{fig:1es0229_b15_ebllow}]{
\includegraphics[width=70mm]{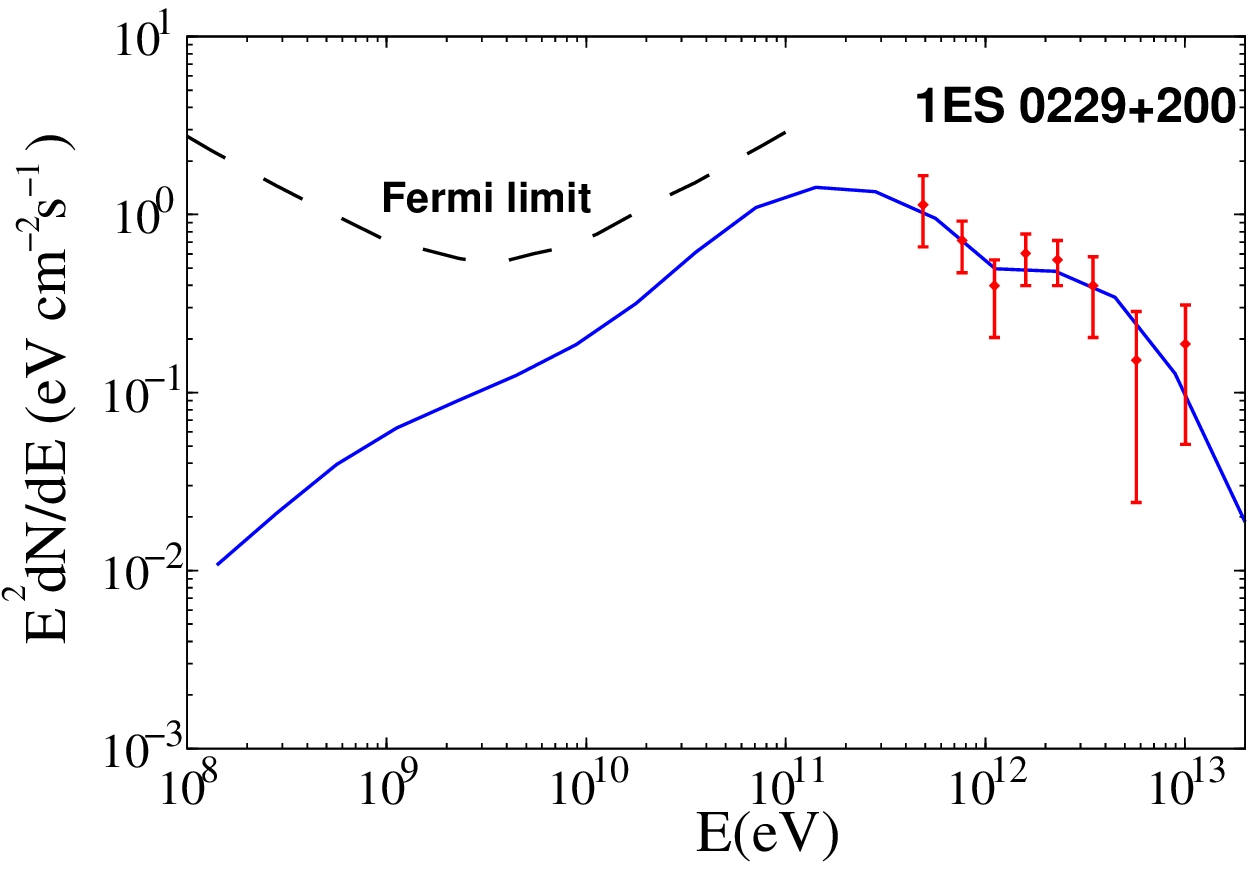}}
& 
\subfigure[\label{fig:1es0229_b15_eblhigh}]{
\includegraphics[width=70mm]{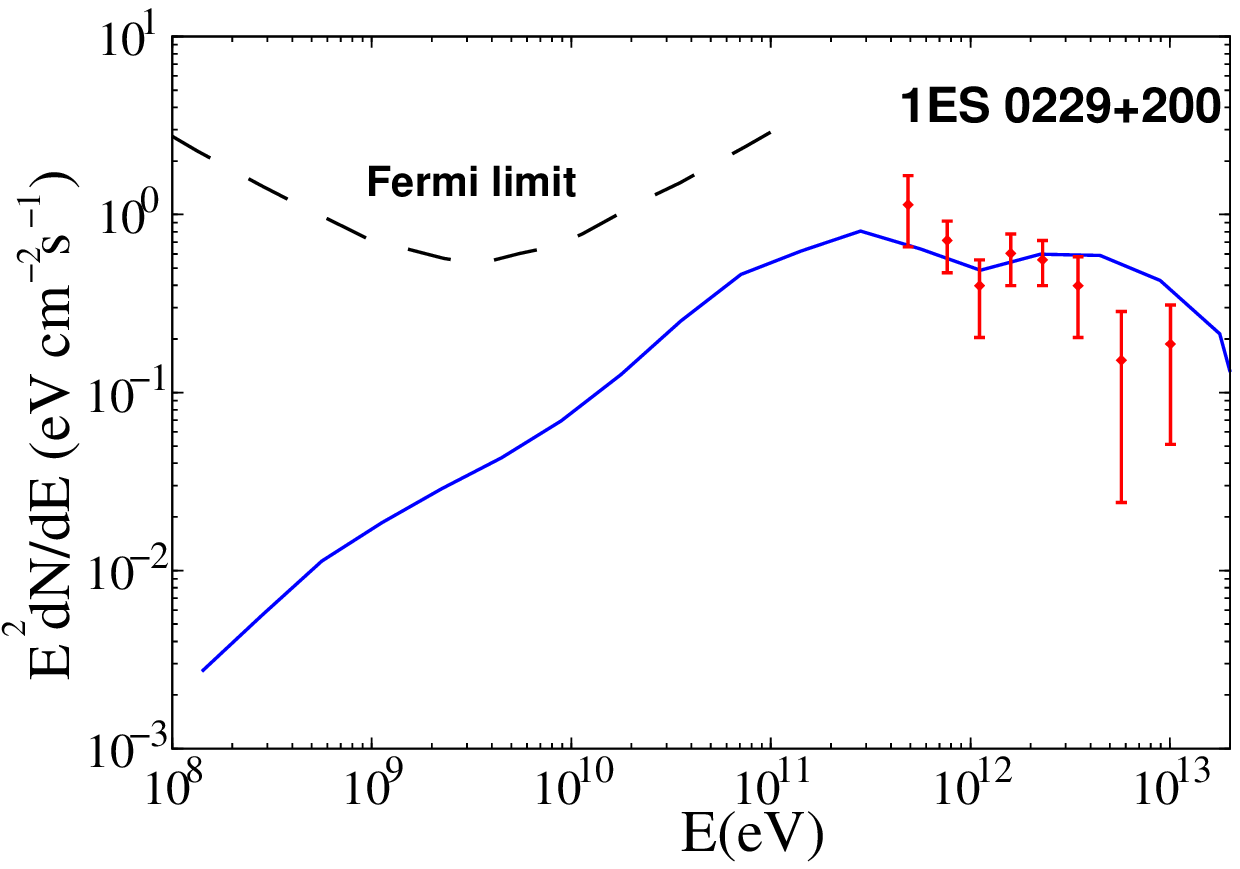}}
\\
\subfigure[\label{fig:1es0347_b15_ebllow}]{
\includegraphics[width=70mm]{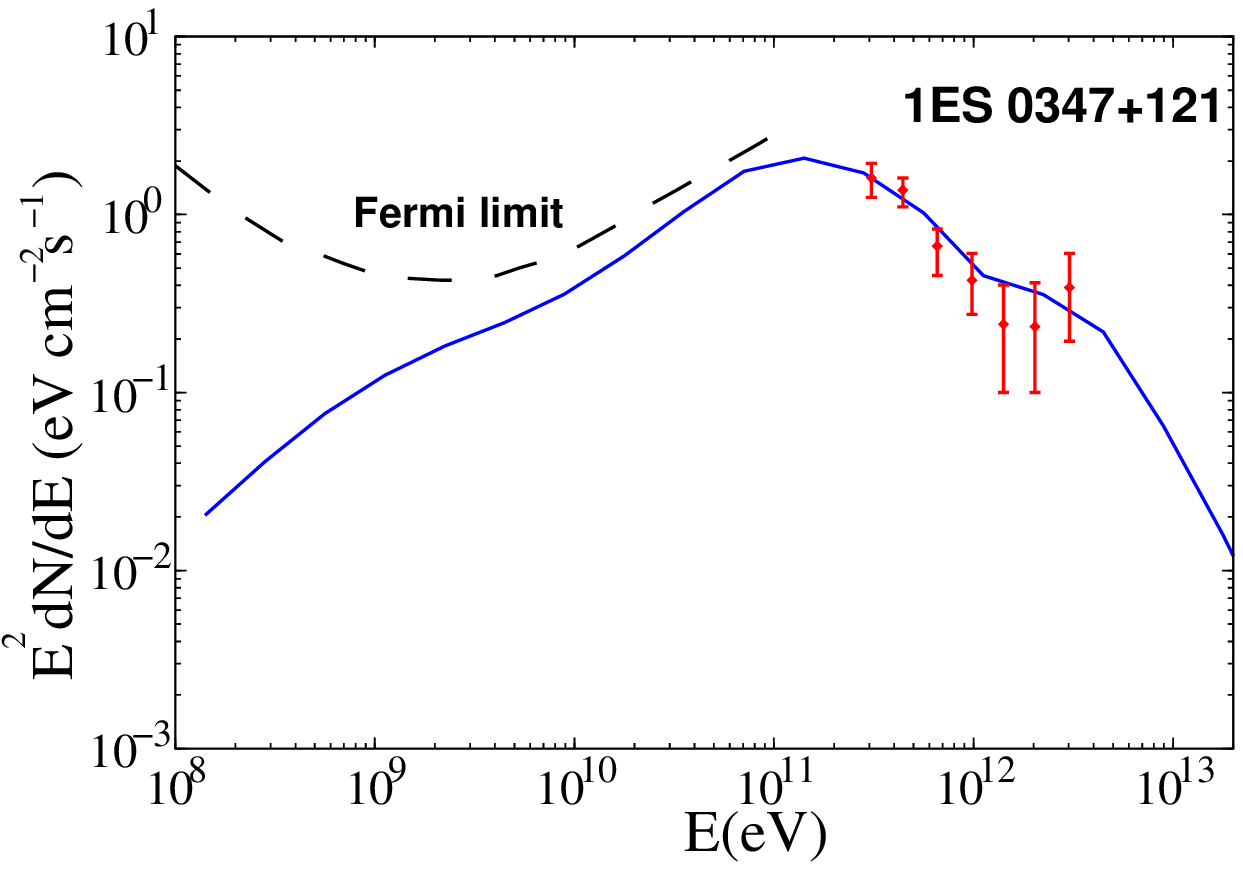}}
& 
\subfigure[\label{fig:1es0347_b15_eblhigh}]{
\includegraphics[width=70mm]{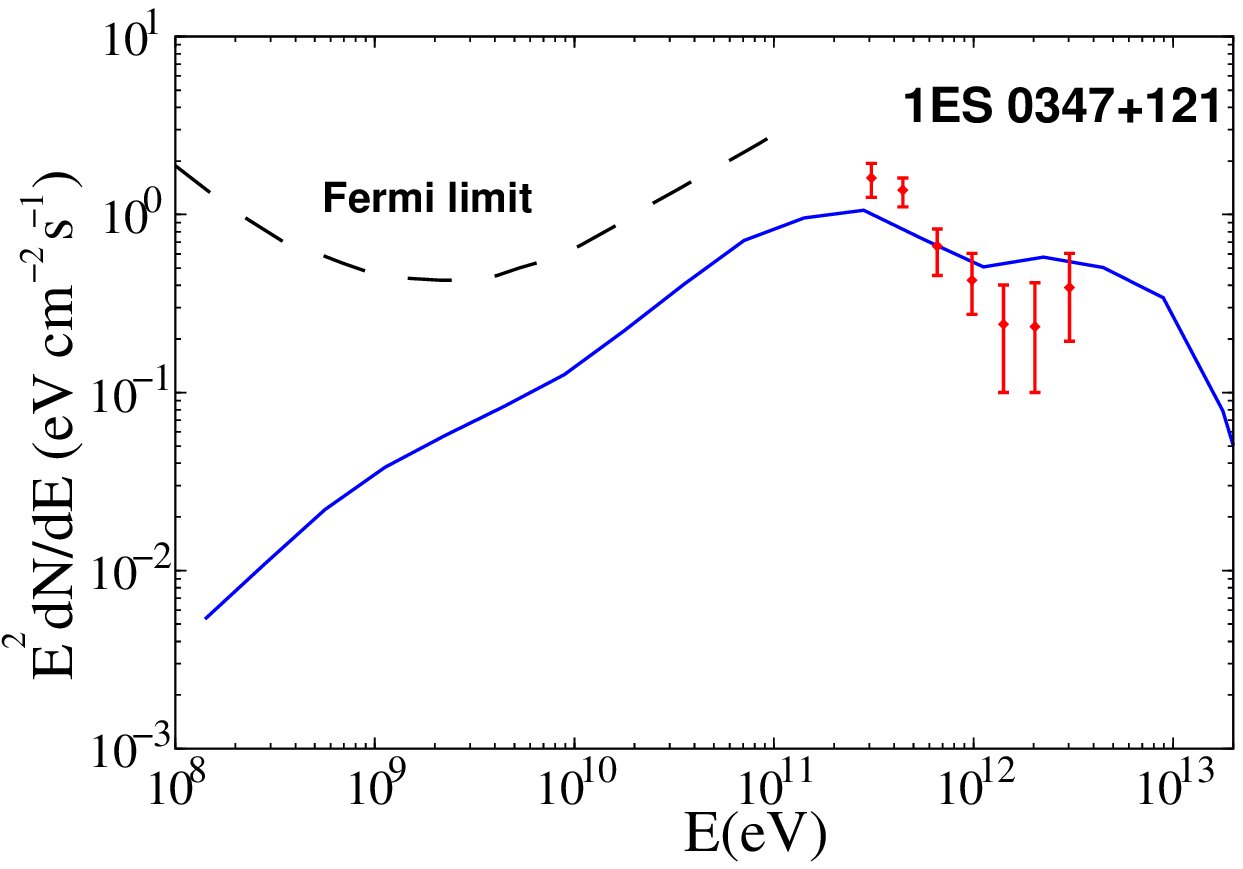}}
\\
\subfigure[\label{fig:1es1101_b15_ebllow}]{
\includegraphics[width=70mm]{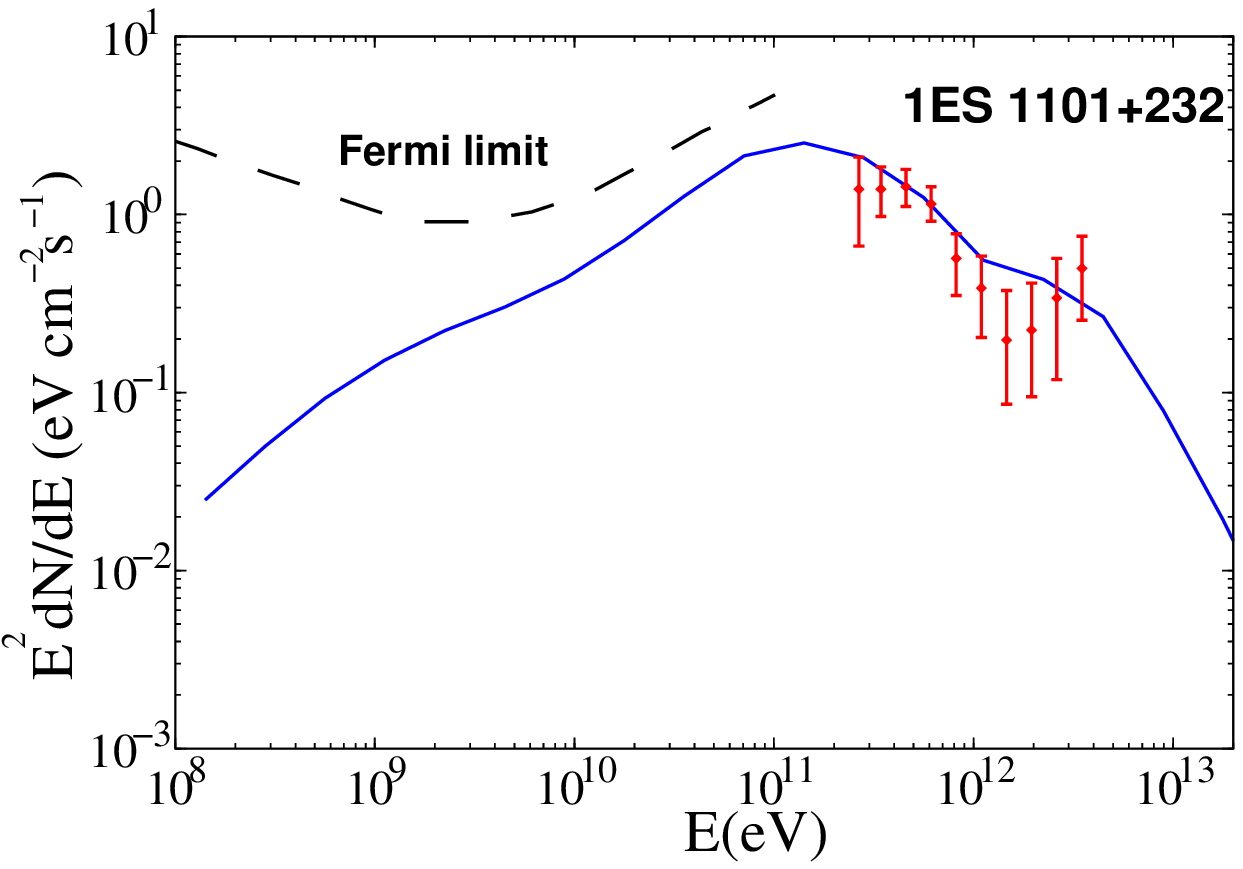}}
& 
\subfigure[\label{fig:1es1101_b15_eblhigh}]{
\includegraphics[width=70mm]{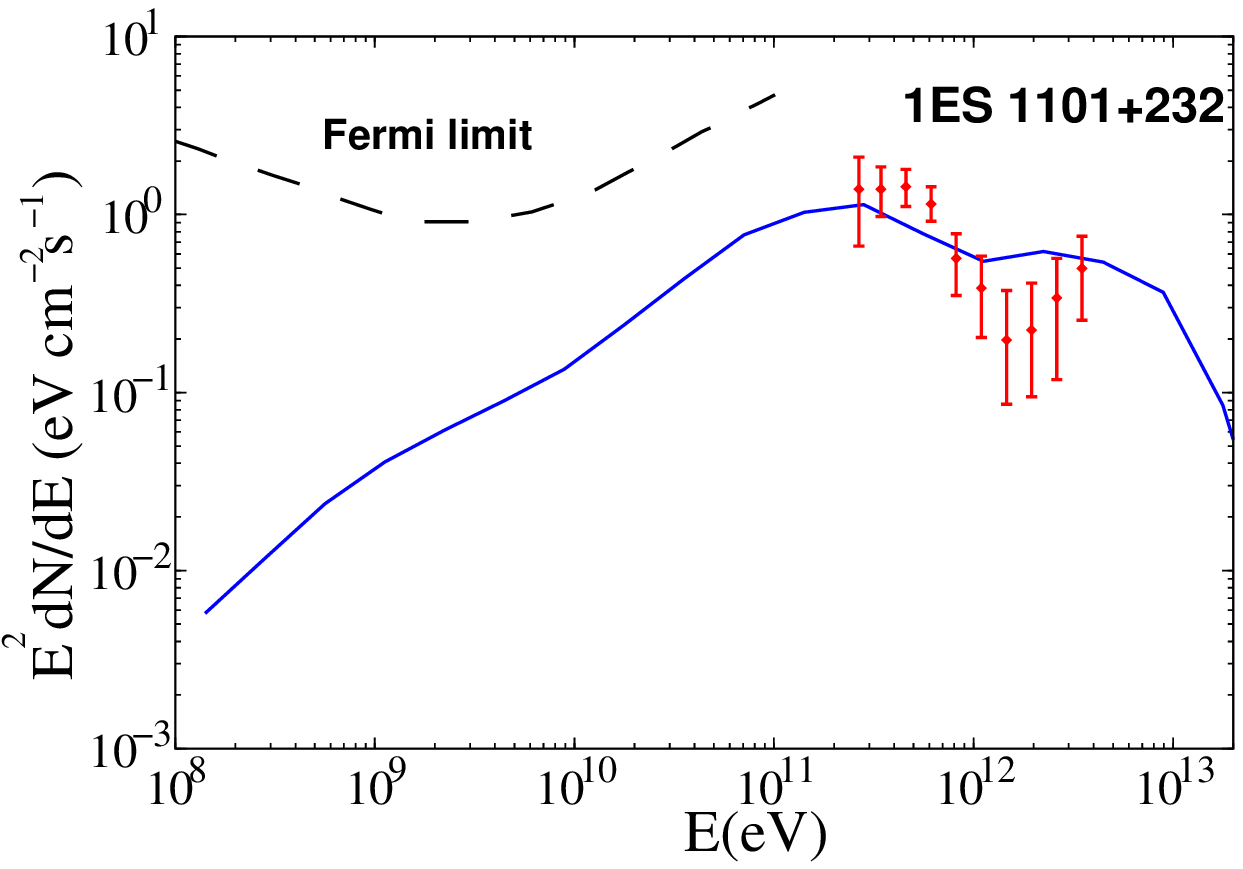}}
\end{tabular}
\caption{
Comparison of the predicted spectra with the HESS data for three blazars: panels (a) and (b) show model prediction and the data for 1ES 0229+200~\citep{2007A&A...475L...9A};
panels (c) and (d)) show the predicted spectrum and the data for 1ES 0347-121~\citep{2007A&A...473L..25A};  panels (e) and (f) show the model prediction and the data for  
1ES 1101-232~\citep{2007A&A...470..475A}. The Fermi upper limits shown at lower energy were derived from the data by~\cite{2010Sci...328...73N}. 
Panels on the left show the prediction for  ``high''  EBL, while panels on the right show the prediction for the ``low'' EBL.  The``high'' EBL is from the model of \cite{2006ApJ...648..774S}, while the 
``low'' EBL is the result of scaling down of ``high'' EBL to the level of 40\%.  (This range encompasses all published models.)  
}
\end{center}
\label{fig:photon_spectra}
\end{figure}

The results for the spectra are presented in Fig.~1.  We have chosen three most distant blazars observed in the
TeV energy range, which show no variability, and which have upper limits set by Fermi: 1ES 0229+200, 1ES 0347-121, and 1ES 1101-232.  We fit the spectra with secondary gamma rays  produced by cosmic-ray interactions along the line of sight.

\begin{figure}
\begin{center}
\includegraphics[width=0.8 \textwidth]{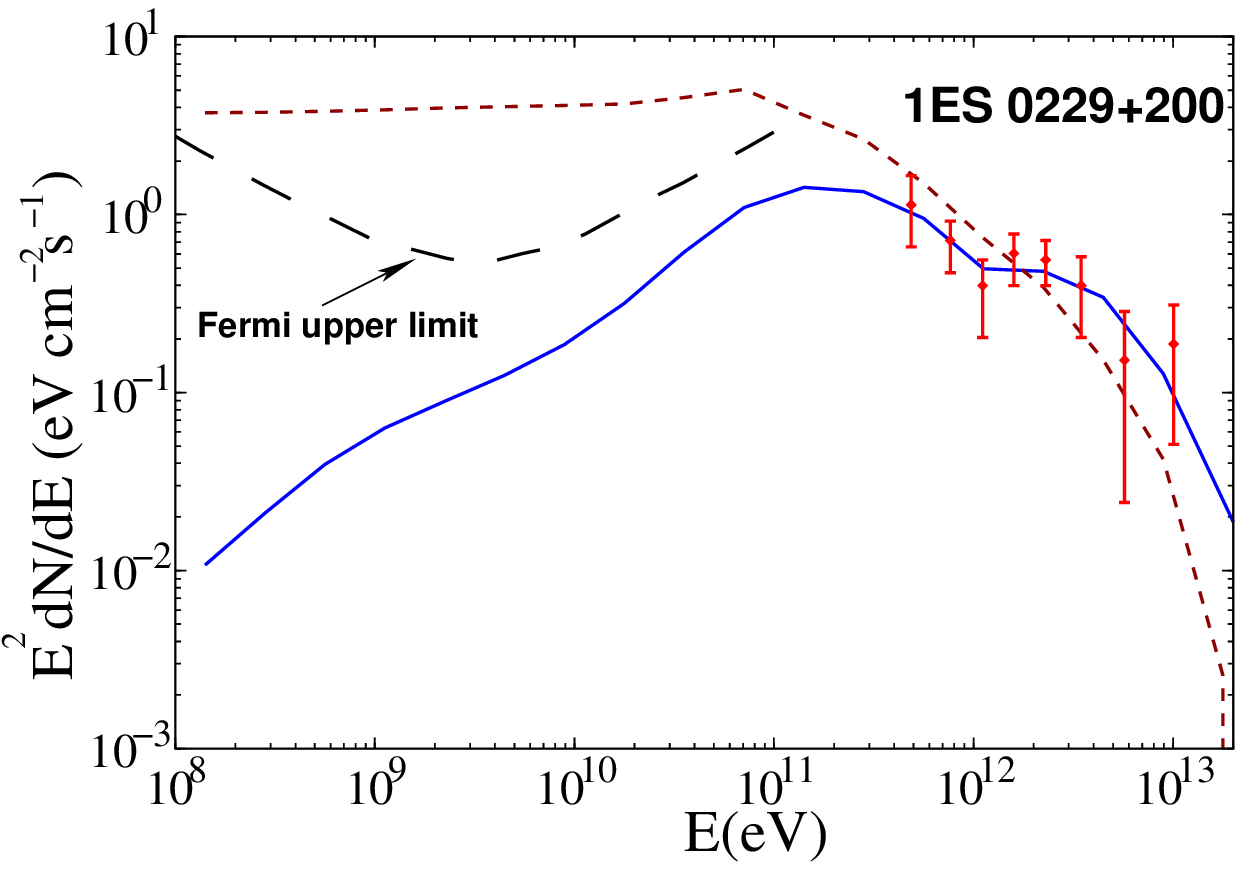}
\caption{
Same as in Fig.~\ref{fig:1es0229_b15_eblhigh}, with the addition of a gamma-ray signal expected in the absence of line-of-sight cosmic-ray contribution, shown by the short-dashed line.  
The intrinsic gamma-ray spectrum was chosen to be a single power law with spectral index $2$.  For this illustrative example, the predicted signal at lower energies 
violates the Fermi bound, but a harder spectrum and/or deviations from a single power law can improve the fit to the data. While an acceptable fit can be obtained without a cosmic-ray contribution,  the secondary gamma rays from cosmic rays fit the data for a broad range of parameters without tuning.  A broad range of spectral indices and magnetic fields, with and without cosmic rays, was considered by \cite{Essey:2010nd}. 
}
\end{center}
\label{fig:no_CR}
\end{figure}

There are several uncertain parameters relevant to this fit.  For the IGMF we chose the average value of $B=1$~fG, coherent over a correlation length $\sim 1$~Mpc; this is consistent with the observations~\citep{Ando:2010rb}. (Effects of the magnetic fields on the goodness of fit will be explored elsewhere.)  The jet opening for the protons was assumed to be $\theta_{\rm jet} = 6^\circ$, which corresponds to a moderate Lorentz factor of $\Gamma=10$. The choice of $\theta_{\rm jet}$ will only have an effect on the spectrum if the deflections of the secondaries are greater than $\theta_{\rm jet}$. For the magnetic fields considered in this paper this is only true for gamma rays in the GeV range. Thus the fits to the TeV energy range are insensitive to the choice of $\theta_{\rm jet}$. However, the choice of $\theta_{\rm jet}$ does affect the overall luminosity normalization, as shown in Fig.~5 and discussed below. The spectrum of protons emitted by the AGN is unknown, but as shown by~\cite{Essey:2009ju}, the spectra of secondary photons are not sensitive to the spectral index $\alpha$ and the maximal energy $E_{\rm max}$ of the protons in a broad range of parameters consistent with the data on UHECR~\citep{Berezinsky:2002nc}.  We parameterize the proton spectrum by a simple power law: 
\begin{equation} 
 F(E)\propto E^{-\alpha} \ \theta (E_{\rm max}-E).
\end{equation}
We allow $\alpha$ to vary between 2 and 2.7 because $\alpha=2$ and $\alpha=2.7^{+0.05}_{-0.15} $ give a good fit to the cosmic ray data at low energy and at the highest energies, respectively~\citep{Berezinsky:2002nc}. In fact, a satisfactory fit to the data can be obtained for any value of $1.5<\alpha<3$. This should be contrasted with fitting the data in the absence of cosmic ray contribution, in which case very hard intrinsic gamma-ray spectra of distant blazars are required ({\em cf.}~Fig.~\ref{fig:no_CR} and the recent paper of \cite{Essey:2010nd}). 

What is also uncertain is the power produced by AGN in cosmic rays, $L_p$.  We fit the photon data by choosing this unknown parameter.  The predictions shown in Fig.~1 are not sensitive to most of the parameters listed above, except for $L_p$. Furthermore, while the spectra depend on the model of EBL for a fixed value of $L_p$, the uncertainty in EBL is much smaller than the uncertainty in $L_p$, so, to first approximation, it could be absorbed by the uncertainty in $L_p$. For blazars, relativistic beaming effects must also be taken into account when calculating $L_p$.  The isotropic (unbeamed) equivalent luminosity $L_{p,{\rm iso}}$ can be much greater than $L_p$:
\begin{equation}
L_{p, {\rm iso}} = \left( \frac{2}{1-\cos \theta_{\rm jet}} \right) L_{p}. 
\end{equation}
As discussed below, a more detailed goodness-of-fit analysis does show some preference for some models of EBL. To encompass all the published models using a simple parameterization, we have taken the ``high'' EBL model of \cite{2006ApJ...648..774S} as the upper limit and an EBL spectrum of the same shape but scaled down to the level of 40\% as the lower limit.  A number of models fall in this range~\citep{1998ApJ...493..547S,2002A&A...386....1K,Kneiske:2004,2007ApJ...667L..29S,Franceschini:2008tp,Horiuchi:2008jz,Primack:2008nw,2009MNRAS.399.1694G,2009ApJ...697..483R,2010ApJ...712..238F}.

The model predictions presented in Fig.~1 are essentially one-parameter fits, with the shape of the spectrum fixed by the shape of EBL, and the overall height proportional to the product of $L_p$ and the level of EBL.  
The parameters used for the spectra shown in Fig.~1 and the goodness of each fit are shown in Table~1. 

\begin{deluxetable}{lllllll}  
\tablecolumns{7}
\tablecaption{Model parameters for the spectra shown in Fig.~1. (Here we assumed $E_{\rm max}=10^{11}$~GeV, $\alpha=2$, and $\theta_{jet}=6^\circ$.)}
\tablehead{   
  \colhead{Source} &
  \colhead{Redshift} &
  \colhead{EBL Model } &
  \colhead{ $L_p$ } &
  \colhead{ $L_{p, {\rm iso}}$ } &
  \colhead{$\chi^2$}& 
  \colhead{DOF}
}
\startdata
1ES0229+200 & 0.14 & Low & $1.3 \times 10^{43}$ erg/s & $4.9 \times 10^{45}$ erg/s & 6.4 & 7 \\
1ES0229+200 & 0.14 & High & $3.1 \times 10^{43}$ erg/s & $1.1 \times 10^{46}$ erg/s & 1.8 & 7 \\ 
1ES0347-121 & 0.188 & Low & $2.7 \times 10^{43}$ erg/s & $1.0 \times 10^{46}$ erg/s & 16.1 & 6 \\ 
1ES0347-121 & 0.188 & High & $5.2 \times 10^{43}$ erg/s & $1.9 \times 10^{46}$ erg/s & 3.4 & 6 \\ 
1ES1101-232 & 0.186 & Low & $3.0 \times 10^{43}$ erg/s & $1.1 \times 10^{46}$ erg/s & 16.1 & 9 \\ 
1ES1101-232 & 0.186 & High & $6.3 \times 10^{43}$ erg/s & $2.3 \times 10^{46}$ erg/s & 4.9 & 9 
\enddata
\label{table}
\end{deluxetable}

We have obtained an excellent fit for each of the three blazars using some reasonable values of 
cosmic ray power, consistent with theoretical models of cosmic rays~\citep{Berezinsky:2002nc}.  
Furthermore, we have obtained a reasonably good fit for any model of EBL, although 
the power in cosmic rays required in each case depends on the level of EBL.  If anything, ``high'' EBL models are favored, but it will take more data to achieve a statistically significant discrimination between different EBL models.  This conclusion is different from the conclusions of other authors, who considered only the primary photons to fit the data~\citep{Aharonian:2005gh,Mazin:2007pn,2007A&A...475L...9A,2009ApJ...698.1761F,2010ApJ...723.1082A}.  It is easy to understand the origin of this discrepancy.  If one tries to fit the data with primary photons alone, distant sources should show much more significant degree of absorption at high energies.  The apparent lack of absorption can be compensated by lower EBL and harder injection spectra.  However, for secondary photons, the absorption is much less dramatic because a large number of observed photons are produced relatively close to Earth.   Furthermore, higher EBL density actually increases the production of gamma rays in $p\gamma$ interactions. 

The effect of different $E_{\rm max}$ and $\alpha$ is to change the relative contribution of reactions (\ref{PPP}) and (\ref{photopion:0}) to the flux of secondary gamma rays.  If the proton spectrum extends to very high energies (as suggested by data on ultrahigh-energy cosmic rays), then the PPP reaction on CMB  (\ref{PPP}) dominates over pion photoproduction on EBL (\ref{photopion:0}). If, however, $E_{\rm max}$ is small (but larger than the threshold of $10^7$~GeV), the pion photoproduction on EBL (\ref{photopion:0}) is the dominant source of gamma rays.  The only difference for the gamma-ray spectra is the power $L_p$ required from a given source.  Since this power is unknown, one can obtain equally good fits to the gamma-ray data for different values of $E_{\rm max}$, as shown in Figs.~3 and 4.  The only differences in the photon spectra occur at energies well above the reach of data.  

Of course, the luminosity of the source in protons required to fit the observed flux depends on both  $E_{\rm max}$ and $\alpha$.  This dependence is shown in Fig.~5.

\section{Neutrinos from distant blazars}

Neutrino spectra are much more sensitive to the values of $E_{\rm max}$ and $\alpha$.  This is because the shape of the gamma ray spectrum at the end of the electromagnetic cascade is determined mainly by the spectrum shape of the EBL, regardless of whether the main contribution comes from reaction (\ref{PPP}) or reactions (\ref{photopion:0}) and (\ref{photopion:+}).  The overall normalization can be fit with the product of $L_p$ and the level of EBL.  However, only the reaction  (\ref{photopion:+}) produces neutrinos.  For a larger $E_{\rm max}$ and a harder spectrum, more gamma rays come from proton pair production (\ref{PPP}), while the relative contribution required from pion producing reactions is smaller.  This changes the flux of neutrinos relative to photons, as shown in Figs.~3 and 4. If $E_{\rm max}$ is greater than GZK cutoff, $p\gamma_{\rm CMB}$ interactions of protons with cosmic microwave background radiation contribute to the neutrino flux at the highest energies.  Point sources of high-energy neutrinos have not been observed so far, but IceCube detector may be able to detect such neutrinos~\citep{Essey:2009ju}.

Secondary neutrinos obey the $1/d$ scaling law as discussed in the introduction.  This means that, if secondary neutrinos are detected at all, they should be detected from larger distances than the primary neutrinos~\citep{Essey:2009ju}.   
This opens a question about how large
the diffuse intensity from unresolved distant sources is.  Obviously, it 
should not exceed direct measurements.  Furthermore, considering
a solid-angle bin set by the angular resolution of an experiment,
the contribution from the diffuse background should be much
smaller than that from a single source at a typical distance.  This
ensures that the apparent source flux is not modified and that
the source can be detected as a flux excess relative to adjacent
solid-angle bins.  We find that the absolute diffuse intensity from
secondaries is lower than that from primaries, which automatically
ensures these latter requirements.

For a simple estimate, we assume a constant density of identical, non-evolving
sources in a Euclidean universe. 
For primaries, which is the familiar case, the diffuse intensity is
\begin{equation}
dF/d\Omega = \frac{1}{4 \pi} \int_0^R dr\,  L n = \frac{L n R}{4 \pi},
\label{FOmega_primary}
\end{equation}
where $n$ is the number density of sources (including a beaming
factor as necessary), $L$ is the luminosity per source, and $R$ is
the maximum distance.
For secondaries, this is modified by including a factor $(r p)$
in the integrand, where $p$ is the rate of secondary neutrino production 
per unit length of the proton path: 
\begin{equation}
dF/d\Omega = \frac{1}{4 \pi} \int_0^R dr \, L n\, (r p)
= \frac{L n R}{4 \pi} \frac{R p}{2},
\label{FOmega_secondary}
\end{equation}
%
Thus, even thought the diffuse intensity
builds up more rapidly with maximum radius than the usual case,
it is always smaller in absolute value when the probability of a proton to interact is much smaller than 1, and, 
therefore, $Rp \ll 1$, the limit that we consider.  Therefore, the questions raised above about the diffuse 
flux from secondaries are less important than similar questions for
primaries, and those are addressed by the empirical definition of
our model.  In a more complete treatment, the large-$R$ behavior
would be regulated by terms describing evolution of sources and the expansion of the universe. For beamed sources, the beaming factor $b$ should also be included; this amounts to 
replacing $n$ with $n/b$ in Eqns.~(\ref{FOmega_primary}-\ref{FOmega_secondary}).

One can also estimate the minimal source density $n$ for which the diffuse flux begins to compete with the point sources. 
For an instrument with angular resolution $\phi=1\,^{\circ}$, diffuse flux is given by Eqns.~(\ref{FOmega_primary}-\ref{FOmega_secondary}) with 
$ d \Omega = 2\pi \left (1-\cos(\phi/2)\right )\simeq 2.4\times 10^{-4}$. The flux from a point source is $L/(4\pi d^2)$ for primary or $Lp/(4\pi d)$ for secondary neutrinos. 
For a source at $z=0.14$, or $L\simeq 570\, {\rm Mpc}$, one can take $R=6\, {\rm Gpc}$ (which roughly corresponds to 
$z_{\rm max}=3$).  The requirement that the point source flux dominate over diffuse flux implies a condition on the density $n$ of 
sources, or, equivalently, on the average distance $l=n^{-1/3}$ between the sources: 
\begin{equation}
 l > \left \{ 
\begin{array}{ll}
14 \left (10^3/b \right )^{1/3}\, {\rm Mpc,} & {\rm for\ primary\ neutrinos}\\
8 \left (10^3/b \right )^{1/3} \, {\rm Mpc,}  & {\rm for\ secondary\ neutrinos}.
\end{array} \right .
\end{equation}
This condition is satisfied, in both cases, for the brightest sources.

If one can determine the flavor structure of the neutrinos from point sources, one can use it to further distinguish between primary and secondary neutrinos, at least in the case of the primary neutrinos produced in some environments with a high photon density.  The mean free path for reaction $n\gamma \rightarrow \pi^- p$ can be smaller than the decay length of the neutrons.  Therefore, all the neutrinos would be produced from pion decays and would have the flavor ratios $\nu_e:\nu_\mu:\nu_\tau=1:2:0$. 
Neutrino oscillations alter these ratios mainly by the $\nu_\mu\rightarrow \nu_\tau $ conversion, so that the detected flavors are    $\nu_e:\nu_\mu:\nu_\tau=1:1:1$.  However, the secondary neutrino flux should have a non-negligible contribution from neutron decay, $n\rightarrow p e \bar{\nu}_e$, which should alter the 1:1:1 ratio. Of course, the observed signal could be a combination of both primary and secondary neutrinos. The flavor combination of such a combination should still differ from 1:1:1, but the spectra include an additional component. Thus, we expect the signals plotted below to be the lowest neutrino signal to be expected.

\begin{figure}
\begin{center}

\begin{tabular}{cc}
\subfigure[\label{fig:z0.14Emax1E8}]{
\includegraphics[height=70mm,angle=270]{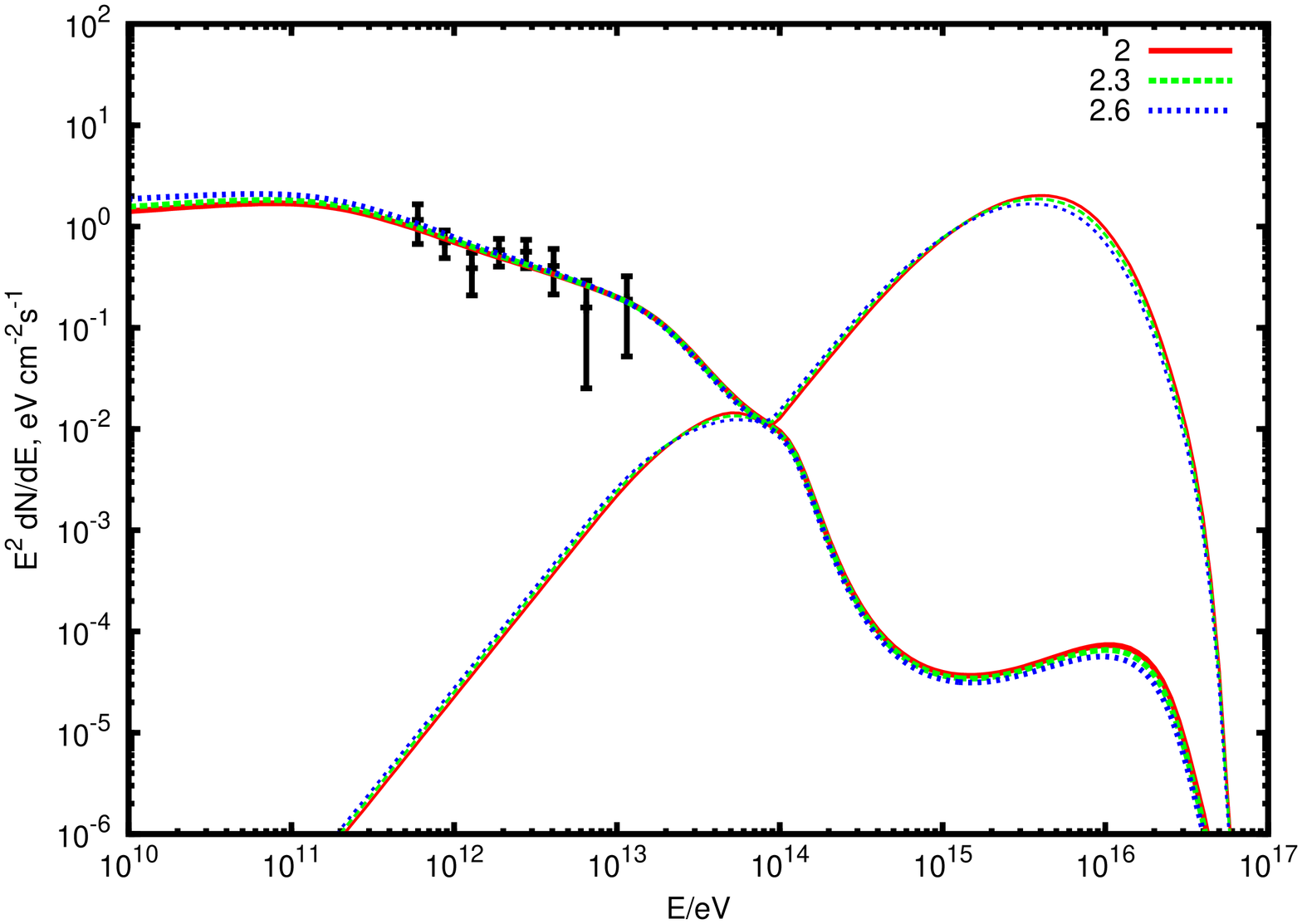}}
& 
\subfigure[\label{fig:z0.44Emax1E8}]{
\includegraphics[height=70mm,angle=270]{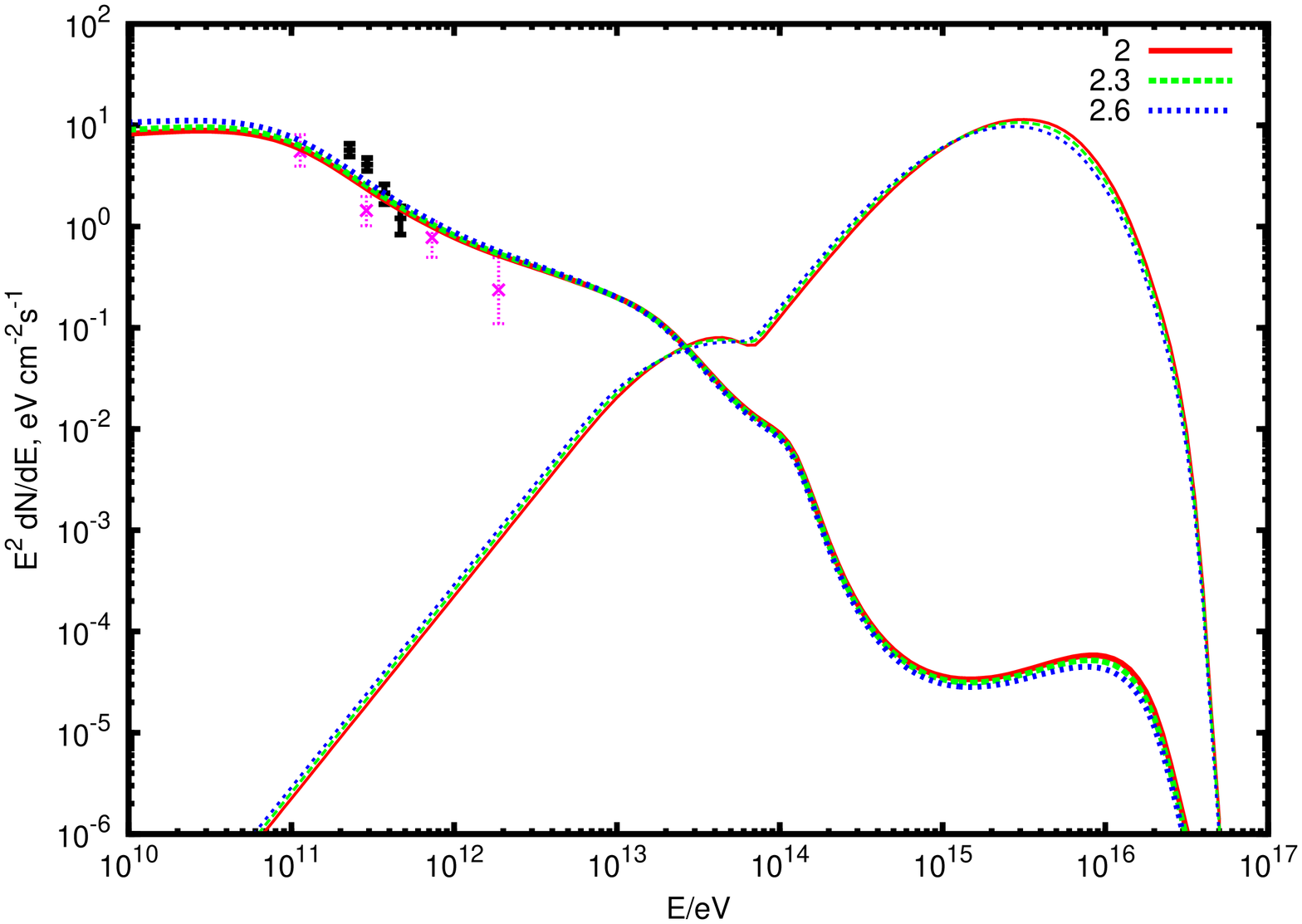}}
\\
\subfigure[\label{fig:z0.14Emax1E10}]{
\includegraphics[height=70mm,angle=270]{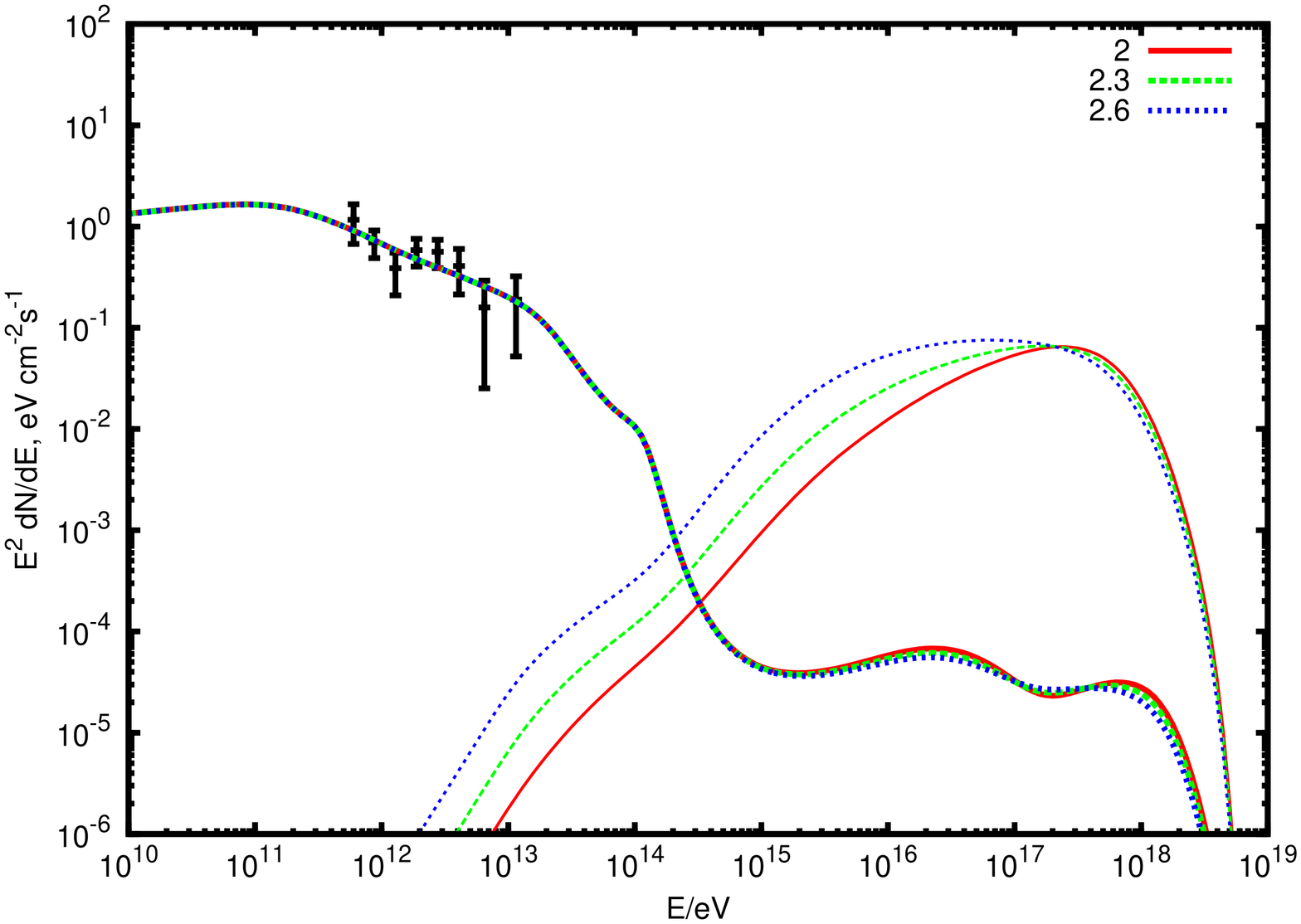}}
& 
\subfigure[\label{fig:z0.44Emax1E10}]{
\includegraphics[height=70mm,angle=270]{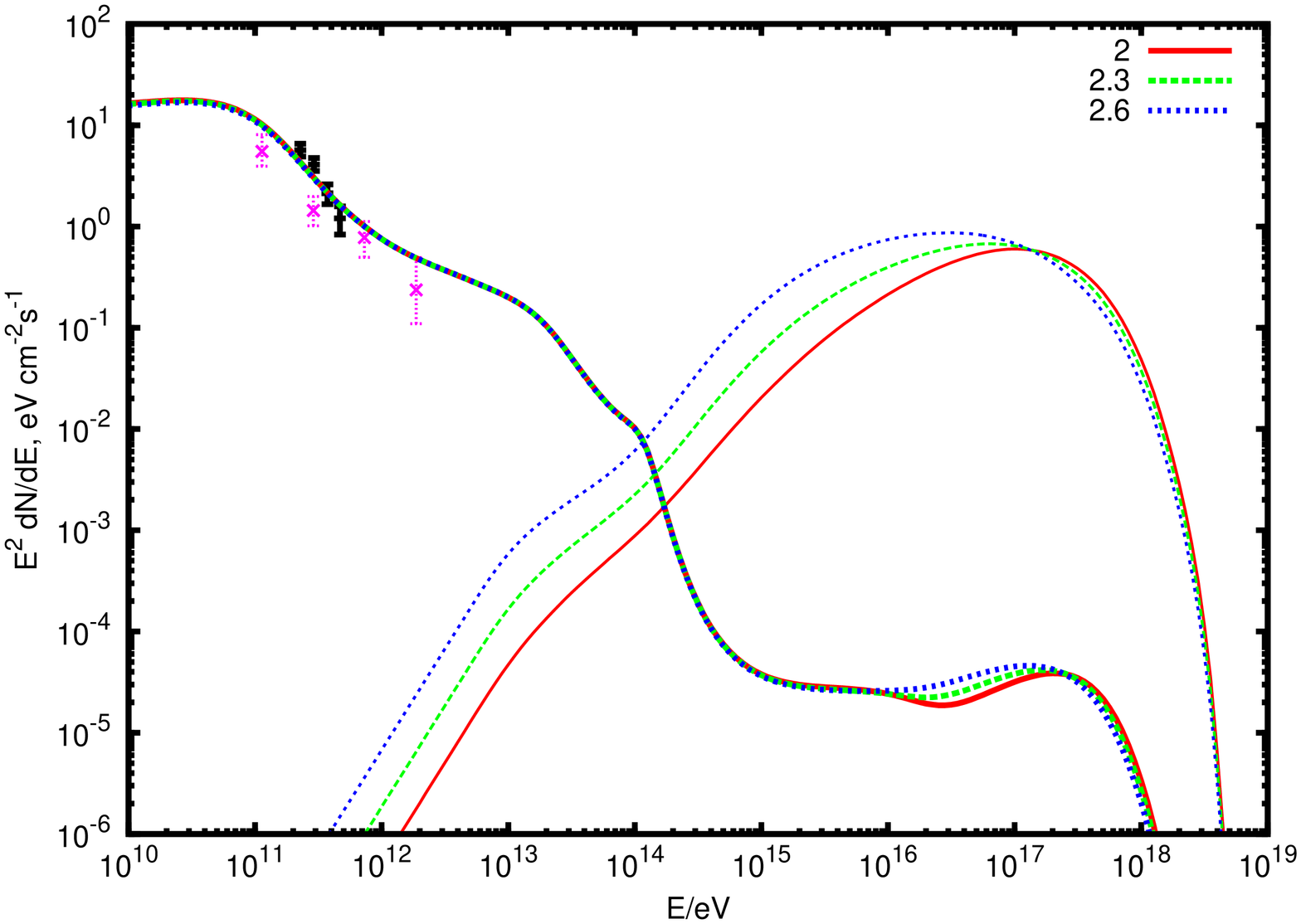}}
\\
\subfigure[\label{fig:z0.14Emax1E11}]{
\includegraphics[height=70mm,angle=270]{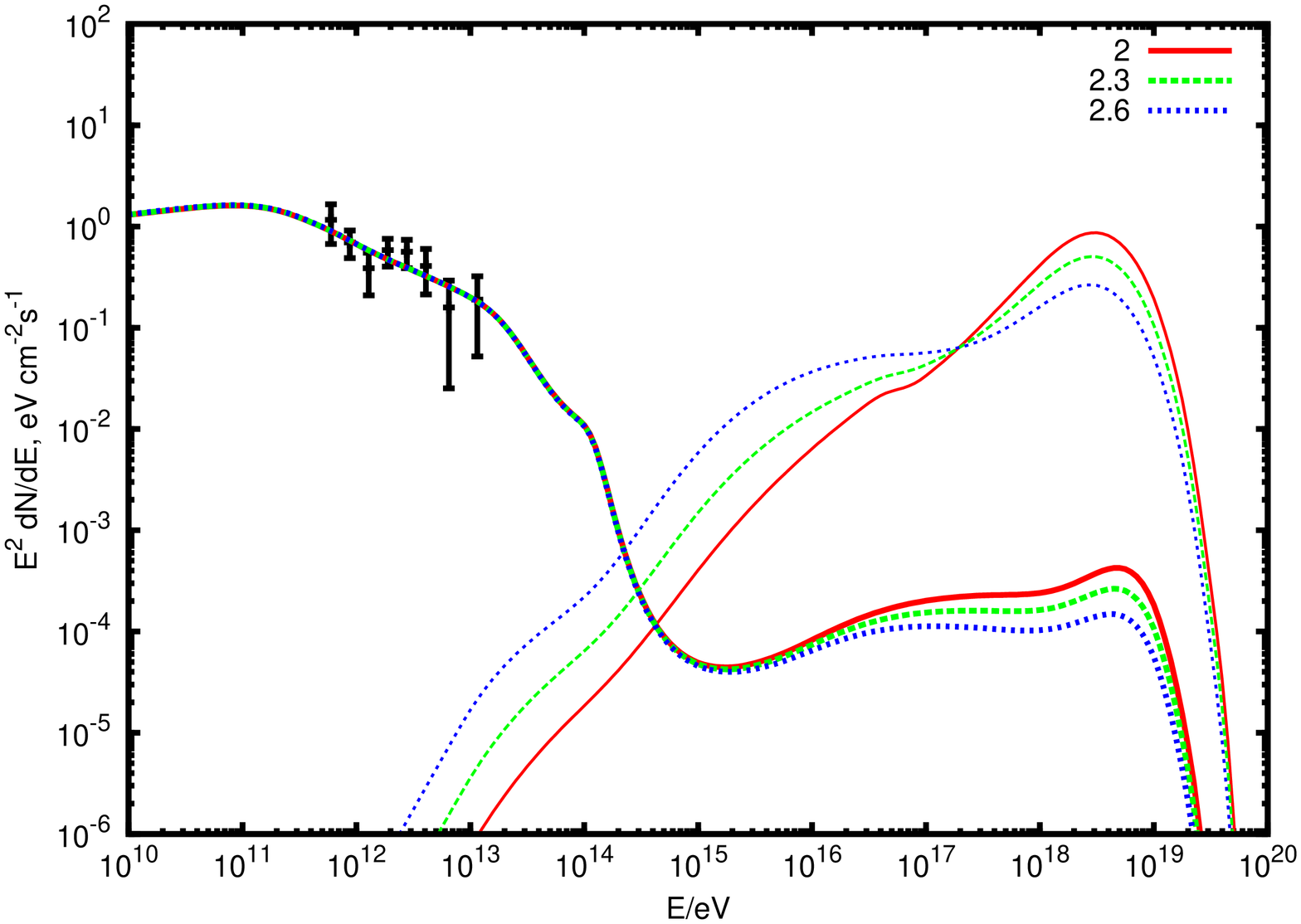}}
& 
\subfigure[\label{fig:z0.44Emax1E11}]{
\includegraphics[height=70mm,angle=270]{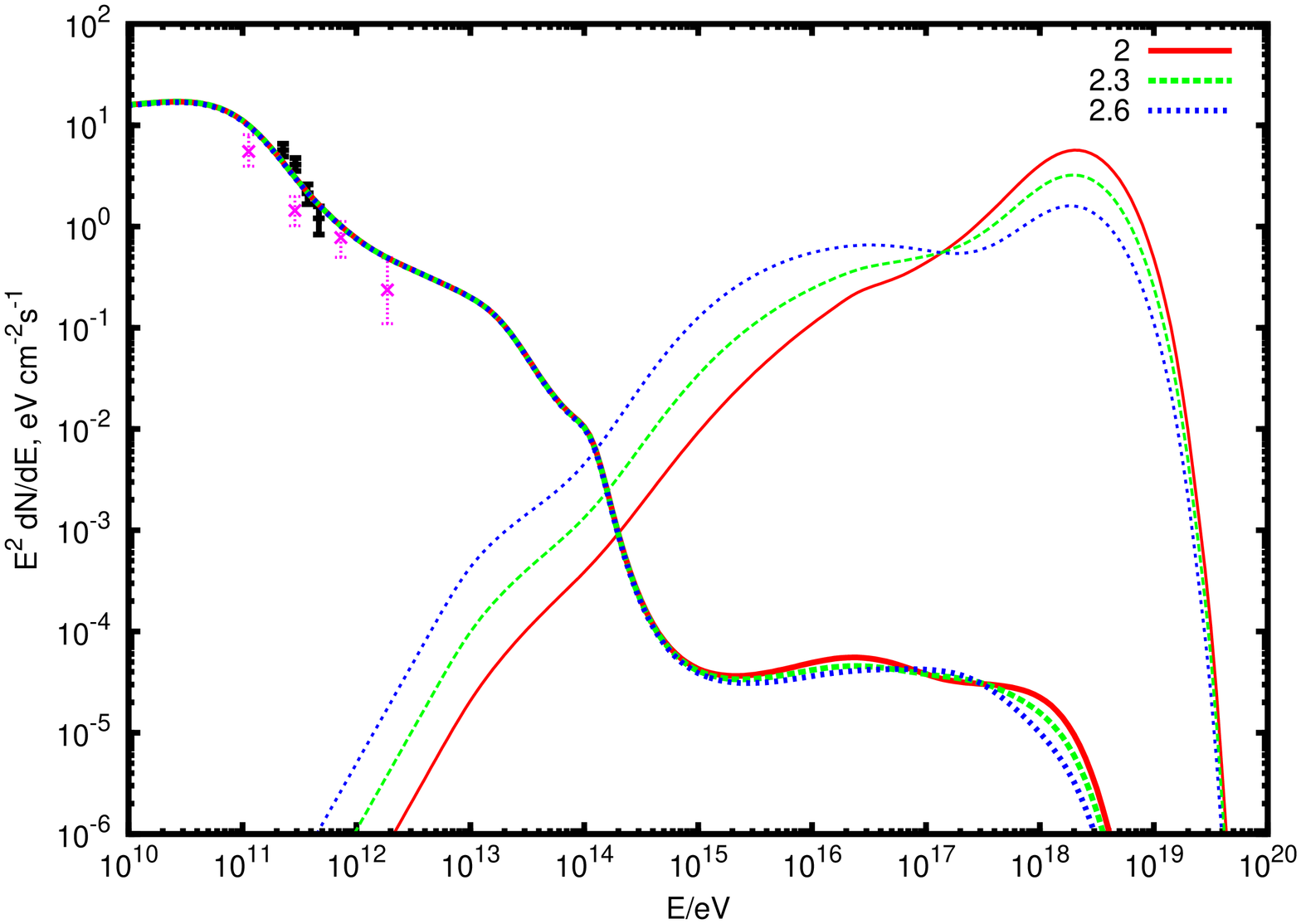}}
\end{tabular}
\caption{
Gamma-ray spectra (lower-energy curves) and neutrino spectra (higher-energy curves)  for different values of maximal proton energy, $E_{\rm max}=10^8 {\rm GeV}, \ 10^{10} {\rm GeV}, \  10^{11} {\rm GeV}$ (top to bottom), for sources at red shifts $z=0.14$ (left), such as blazar 1ES 0229+200, and $z=0.44$ (right), such as blazar 3C66A. The cosmic ray luminosity $L_{\rm p}$ was adjusted to fit the data from  HESS~\citep{2006Sci...314.1424A}, MAGIC~\citep{2008ApJ...685L..23A,2010arXiv1010.0550T}, and VERITAS~\citep{2008ApJ...679..397A,2009ApJ...692L..29A,2009ApJ...693L.104A}. 
Individual curves are labeled by the value of the spectral index $\alpha$.  Here we assume vanishing IGMFs; a $\sim$fG or higher magnetic field would cause some reduction of flux below 1~TeV.
}
\end{center}
\label{fig:phnu1}
\end{figure}

\begin{figure}
\begin{center}

\begin{tabular}{cc}
\subfigure[\label{fig:z0.14alpha2}]{
\includegraphics[height=70mm,angle=270]{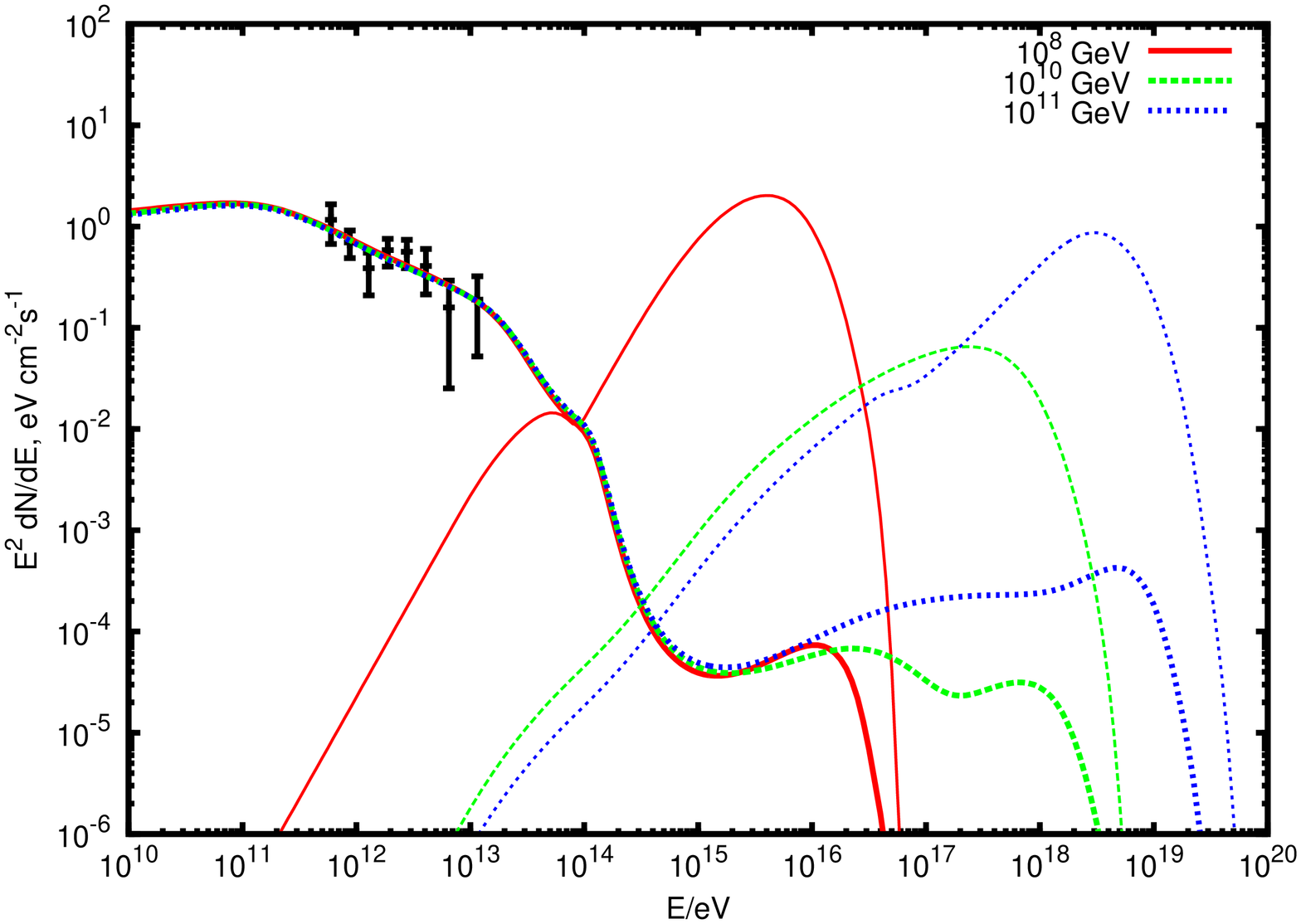}}
& 
\subfigure[\label{fig:z0.44alpha2}]{
\includegraphics[height=70mm,angle=270]{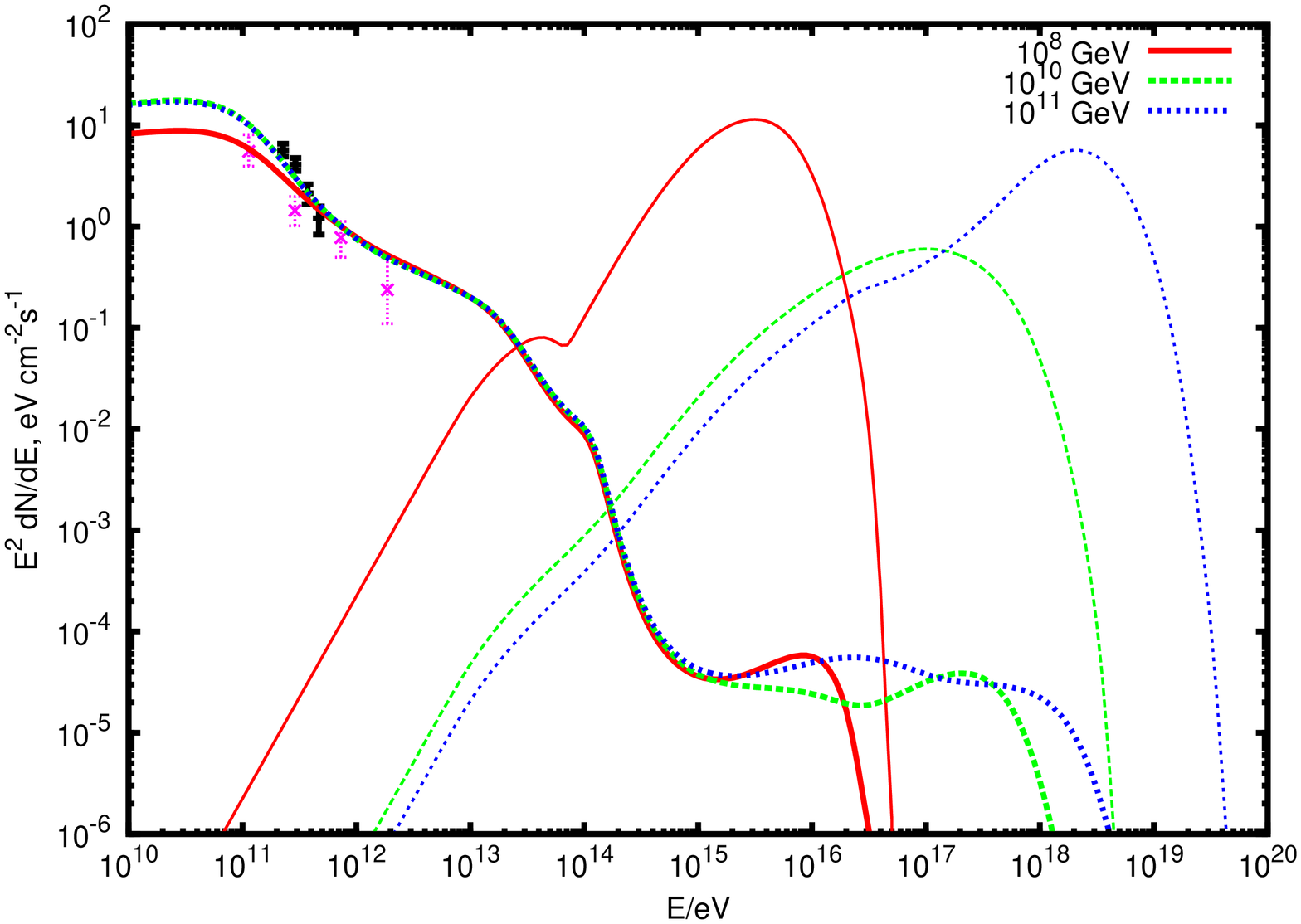}}
\\
\subfigure[\label{fig:z0.14alpha2.3}]{
\includegraphics[height=70mm,angle=270]{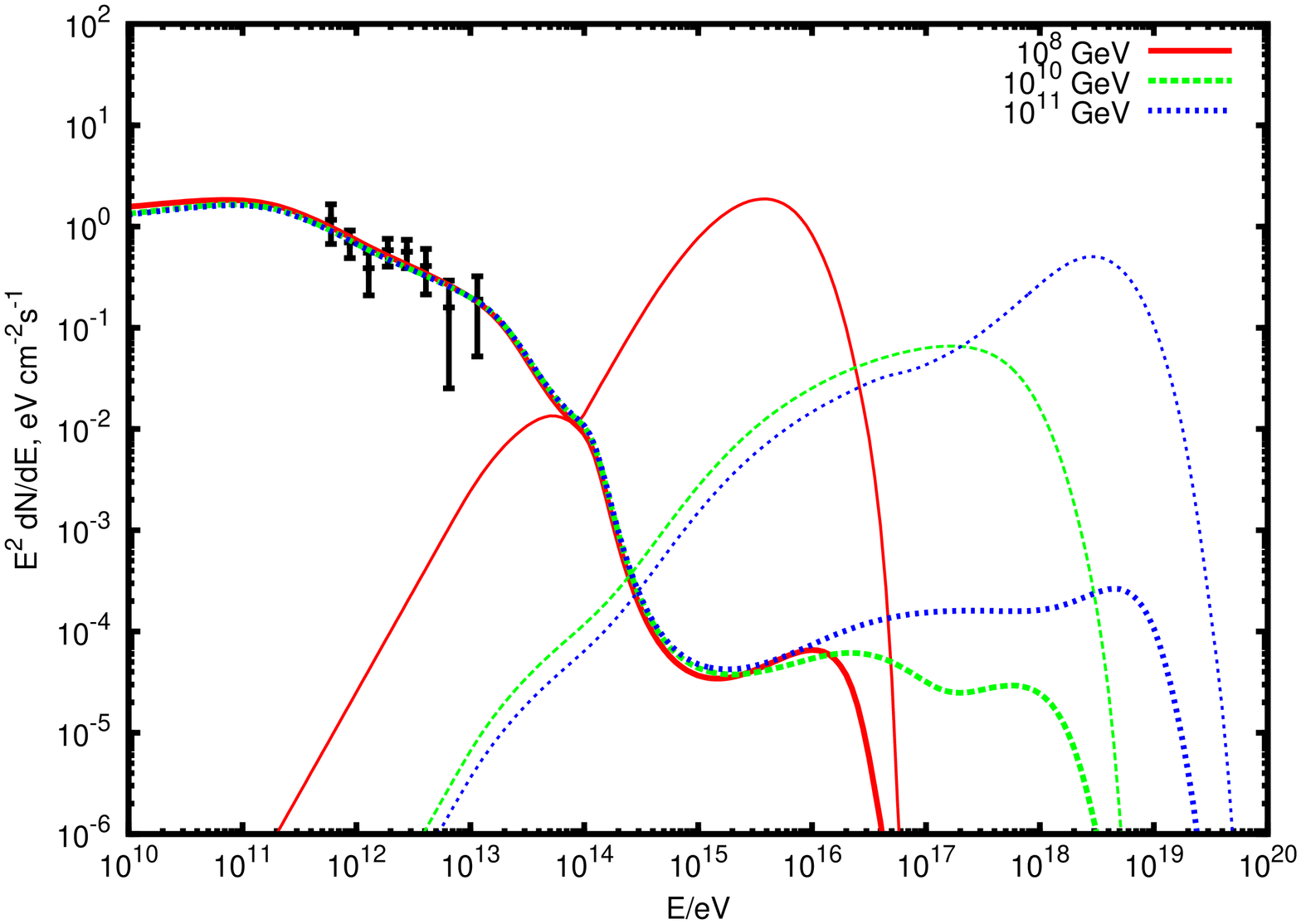}}
& 
\subfigure[\label{fig:z0.44alpha2.3}]{
\includegraphics[height=70mm,angle=270]{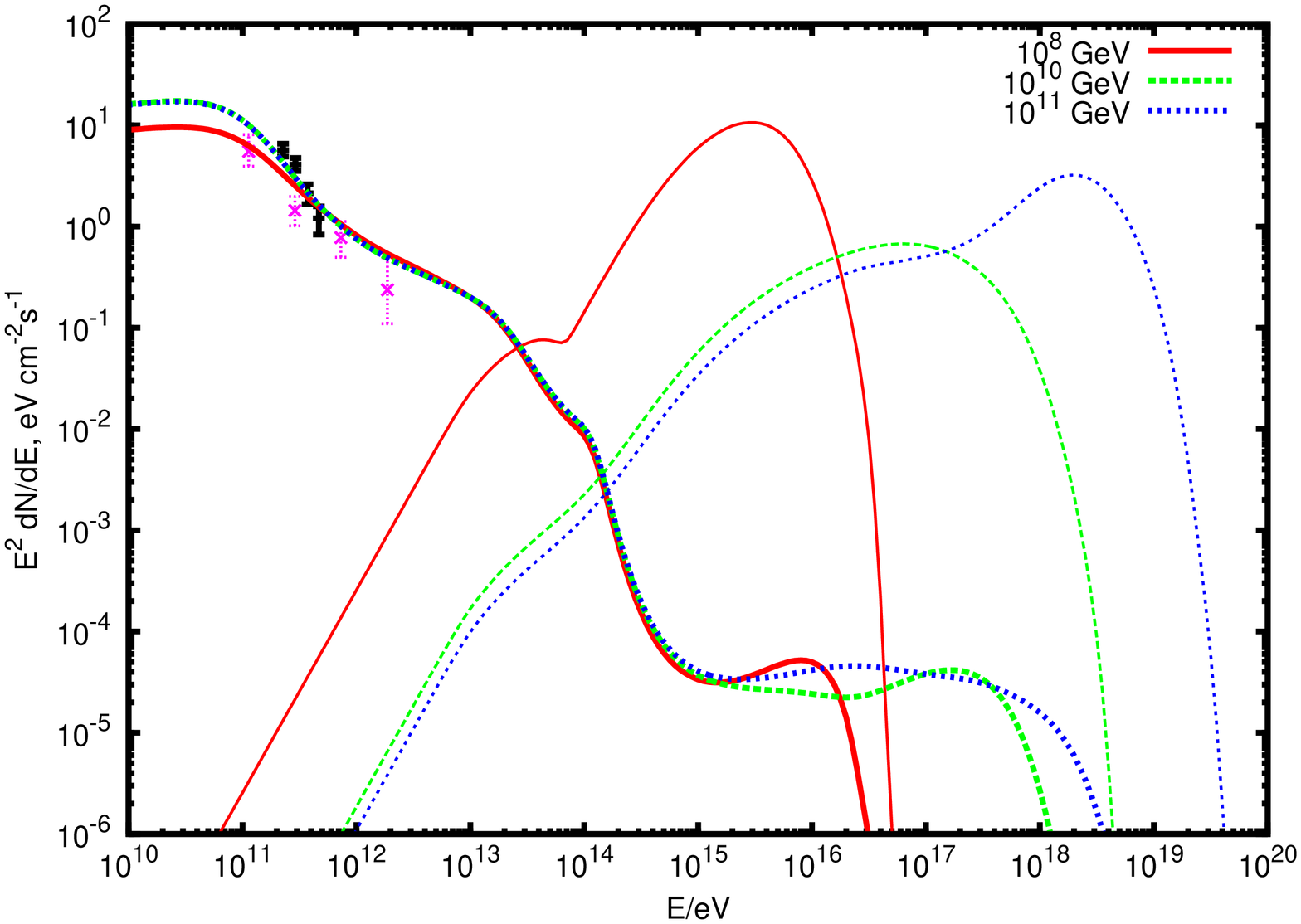}}
\\
\subfigure[\label{fig:z0.14alpha2.6}]{
\includegraphics[height=70mm,angle=270]{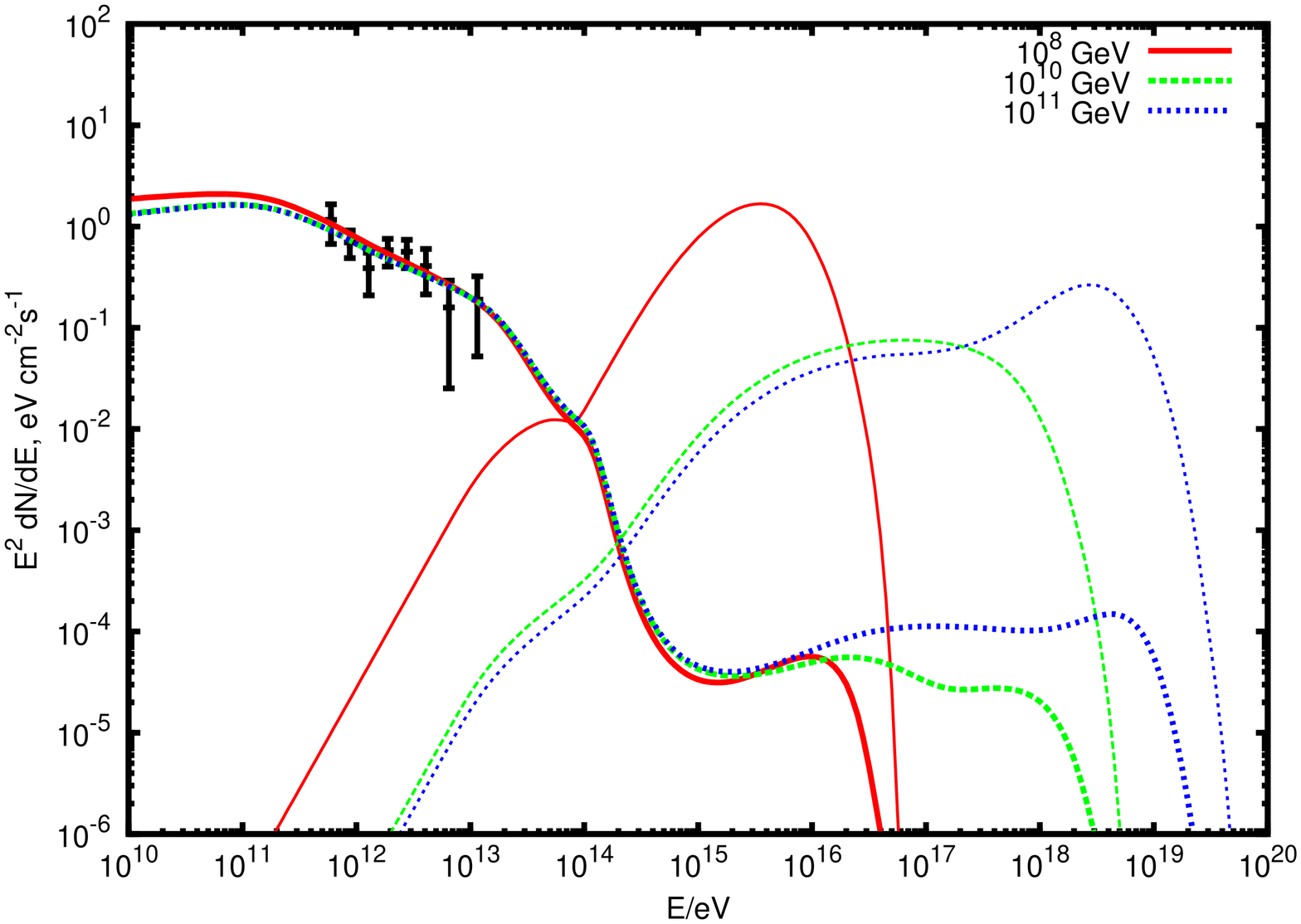}}
& 
\subfigure[\label{fig:z0.44alpha2.6}]{
\includegraphics[height=70mm,angle=270]{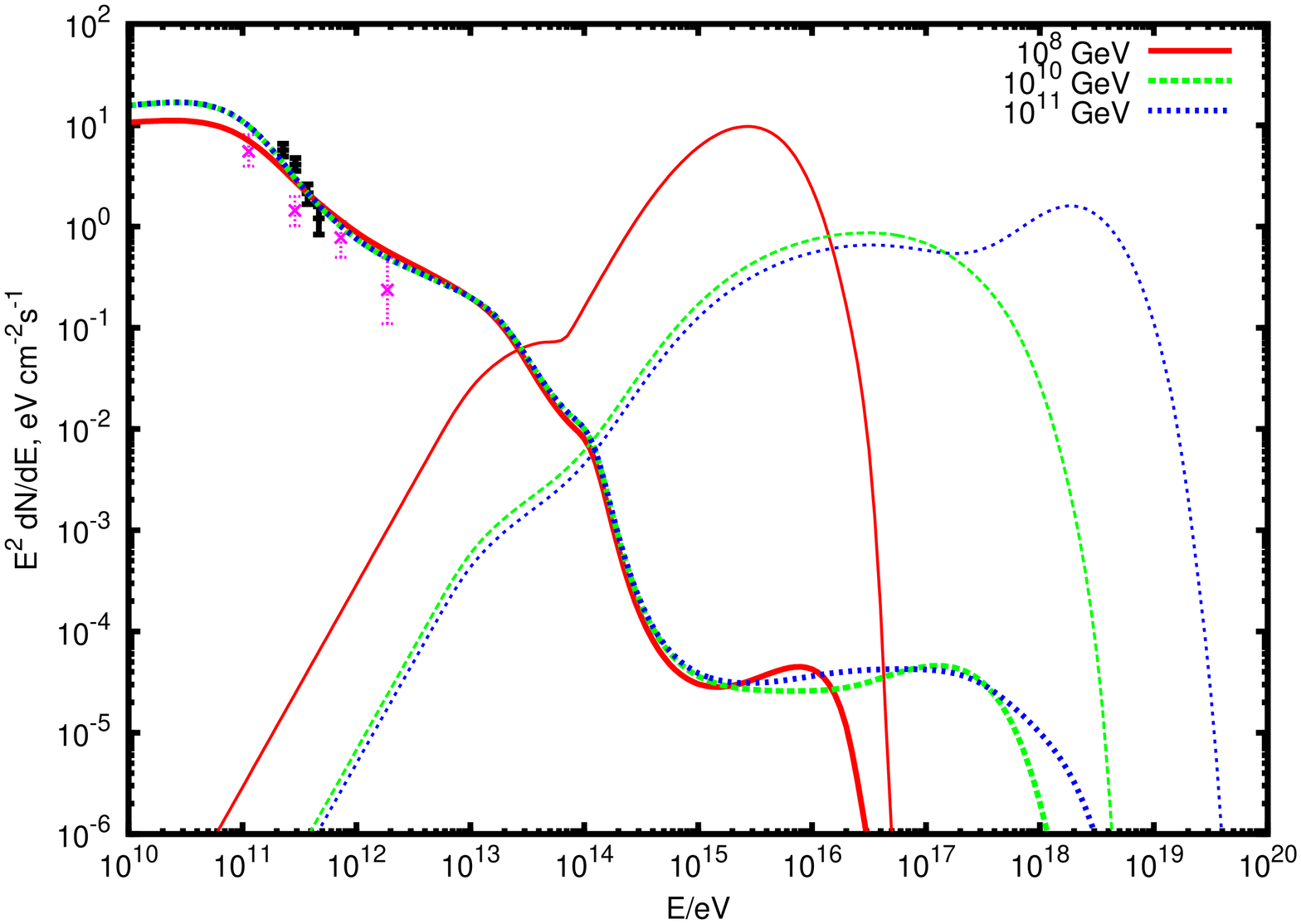}}
\end{tabular}
\caption{
Gamma-ray spectra (lower-energy curves) and neutrino spectra (higher-energy curves)  for $\alpha=2.0, 2.3, \ {\rm and} \ 2.6$ (top to bottom), for sources at red shifts $z=0.14$ (left) and $z=0.44$ (right).   
 The cosmic ray luminosity $L_{\rm p}$ was adjusted to fit the data from  HESS~\citep{2006Sci...314.1424A}, MAGIC~\citep{2008ApJ...685L..23A,2010arXiv1010.0550T}, and VERITAS~\citep{2008ApJ...679..397A,2009ApJ...692L..29A,2009ApJ...693L.104A}. Individual curves are labeled by the value of the maximal proton energy $E_{\rm max}$.  Here we assume vanishing IGMFs; a $\sim$fG or higher magnetic field would cause some reduction of flux below 1~TeV.}
\end{center}
\label{fig:phnu2}
\end{figure}

\section{Implications for extragalactic background light and cosmic ray acceleration models}

Our predictions for the gamma-ray spectra fit the data extremely well for all the models of EBL.  While primary gamma rays are lost to interactions with EBL, secondary gamma rays are produced in these interactions.  The overall flux depends on the EBL photon density multiplied by the unknown and poorly constrained $L_p$, the energy output of AGN in protons. There are small differences in spectral shapes, and one can hope to gain some discriminating power with more data.   At present, the data shows some preference for high EBL, although it is not statistically significant (see Table 1).  (This is in contrast with the limits set on EBL under the assumption that all the observed gamma rays are primary.)  The inclusion of secondary gamma rays brings in one additional 
``free'' parameter $L_p$, but it affects only the overall normalization of the spectrum.  The shape of the spectrum fits the data quite well, much better than the fit one could obtain with the primary sources alone.   Moreover, fitting the data with primary gamma rays demands very hard intrinsic photon spectra, which may be possible~\citep{2007ApJ...667L..29S}, but which are by no means natural or generic, based on a number of theoretical models~\citep{1981MNRAS.196..135P,1987ApJ...315..425K,1988MNRAS.235..997H,1998PhRvL..80.3911B,2001RPPh...64..429M}.  The data thus favor the interpretation in terms of secondary gamma rays, which relaxes the constraints one can put on the models of EBL.  Cosmic ray acceleration models can be improved if the neutrino observations provide insights about the maximal energy and spectral slope.  

In general, secondary gamma-ray flux depends on the present level of EBL more than the EBL evolution, while the neutrino fluxes probe both present and past levels of EBL.  The detection of point sources by IceCube, combined with improved gamma-ray data, can help improve the bounds on both the present level of the EBL and on its evolution.

In principle, one can use gamma-ray data to set upper and lower limits on cosmic ray production in AGN by comparing the primary gamma-ray component with the component generated by cosmic-ray interactions 
along the line of sight.   There is no doubt that the former dominates the signals from nearby blazars, while the latter can take over at 
large distances, where the former is filtered out by photon-photon interactions.  However, since the predicted gamma-ray spectra show 
little sensitivity to the values of the lower and upper energy cutoffs and the spectral slope for cosmic rays, the ratio of the power in 
cosmic rays to the power in gamma rays can vary dramatically, depending on these uncertain parameters.  We have not been able to set 
meaningful limits based on the present data.   Uncertain as they are, the limits on the power in cosmic rays derived from comparison of 
models with cosmic ray data~\citep{Berezinsky:2002nc} appear to be more constraining than any limits one could derive from the present gamma-ray data.

\begin{figure}
\begin{center}
\begin{tabular}{cc}
\includegraphics[width=0.47 \textwidth]{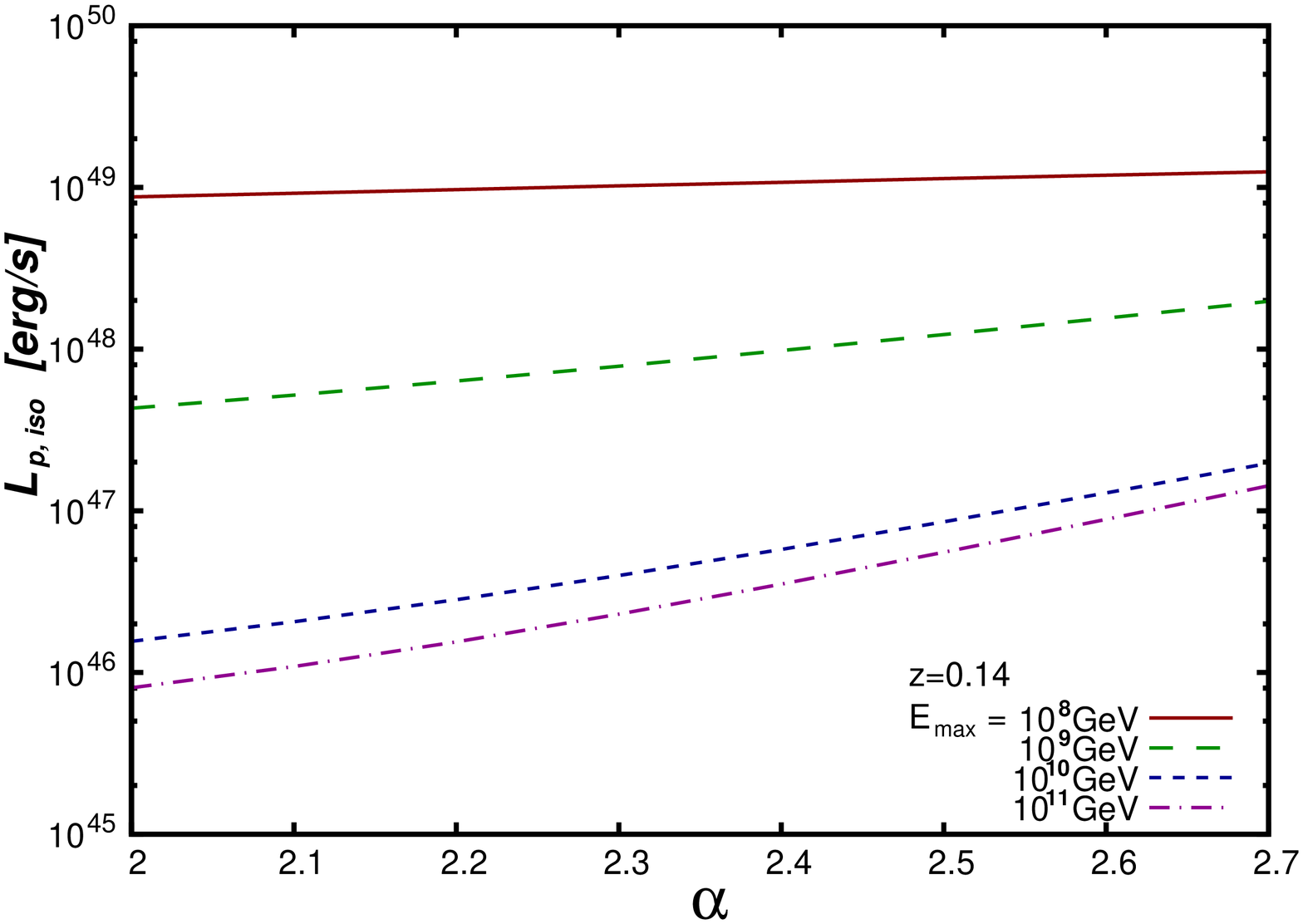}

&
\includegraphics[width=0.47 \textwidth]{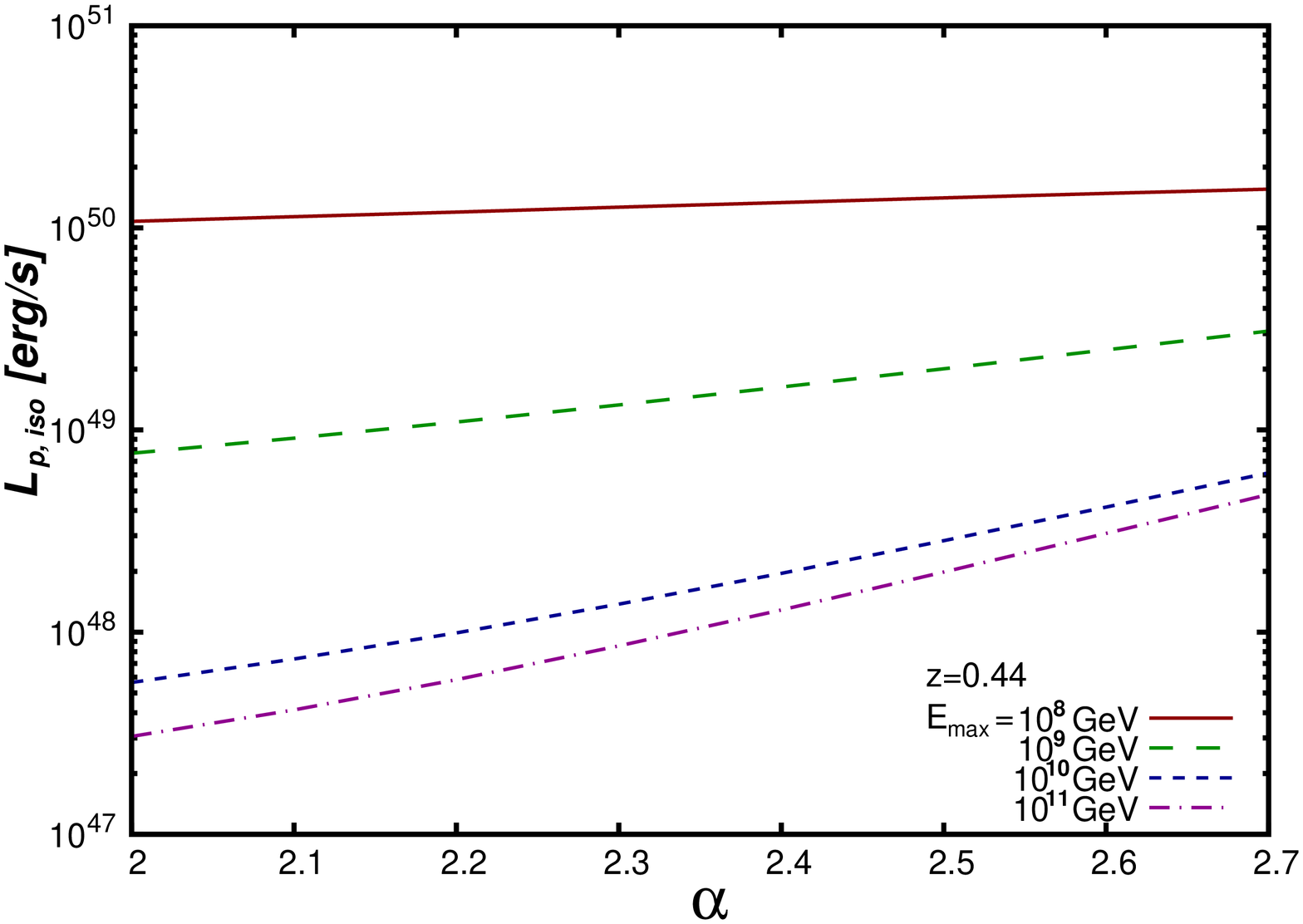}
\end{tabular}
\caption{Dependence of the isotropic equivalent of source power $L_{p, \ {\rm iso}}$ in cosmic rays on the proton spectral index $\alpha$ and the maximal proton energy $E_{\rm max}$ for sources at two redshifts, 1ES0229+200 at $z=0.14$  (left) and 3C 66A at $z=0.44$ (right).  Here IGMF effects are neglected (which results in a small difference with the values given in Table~1).
}
\end{center}
\label{fig:alphaEmax}
\end{figure}

\section{Conclusion}
The surprisingly low attenuation of high-energy gamma rays can be explained by secondary gamma rays produced in interactions of cosmic-ray protons with background photons in the intergalactic space. 
We have obtained excellent fits to observed spectra of several distant blazars, hence extending our prior published results~\citep{Essey:2009zg,Essey:2009ju}. 
All these spectra have a characteristic shape that derives from the {\em shape} of the EBL, and they are less sensitive to the {\em level} of EBL than the spectra of primary gamma rays. 
 At low energies the spectra are harder than predicted by theoretical models \citep{2001RPPh...64..429M,2007ApJ...667L..29S}, which explains why Fermi has not detected these high-energy sources. A future  detection at low energies could help differentiate between the primary and secondary gamma rays. 
Secondary gamma rays are expected to show no temporal variability,  which so far is consistent with the data at energies above TeV. The temporal information is a strong discriminant between primary and secondary gamma rays;  most primary gamma-ray models predict temporal variability on a time scale smaller than the observation time.

We have also presented our predictions for secondary neutrino signals from blazars. The power for the neutrino signal peaks at about $1-10~\rm eV~cm^{-2}s^{-1}$, depending on the choice of source and model parameters. For the declinations of the sources we considered, IceCube sensitivity is, roughly, $10~\rm eV~cm^{-2}s^{-1}$ for 22 strings after 0.75 years and $2~\rm eV~cm^{-2}s^{-1}$  for 80 strings after 1 year \citep{2009ApJ...701L..47A}. This makes the prospects for detection seem plausible.  However, the IceCube sensitivity is calculated assuming a standard $E^{-2}$ spectrum in the energy range from 3~TeV to 3~PeV. A detailed analysis of IceCube sensitivity for the specific spectrum shape predicted for secondary neutrinos and the higher energy range is beyond the scope of this paper.

One can make additional predictions for neutrinos, besides the spectral shape. First, as for gamma rays, there should be no temporal variability observed for neutrinos. Second, the luminosity of sources should vary with distance as $1/d$, as opposed to the usual 1/$\rm d^{2}$ scaling law. This makes more distant sources observable, as compared to predictions for primary neutrinos~\citep{Stecker:1991vm}, and it allows one to confirm the secondary neutrino observations by studying a population of distant sources. Finally, the flavor structure of the observed signal should differ from primary neutrino models due to a significant contribution from neutron decays.

Secondary gamma rays and neutrinos present a new powerful method to probe the radiation and magnetic field contents of intergalactic space, as well as AGN properties. 
The signals differ from primary signals in spectral shape, temporal variability, and scaling with distance. Future ACT and neutrino experiments should consider the effect of specific scaling laws for secondary gamma rays and neutrinos, which bring more sources into the field of view of a given instrument.  

\section{Acknowledgments} 
The authors thank F.~Aharonian, S.~Ando, C.~Dermer, and S.~Razzaque for helpful discussions and comments. The work of W.E. and A.K. was supported  by DOE grant DE-FG03-91ER40662 and NASA ATP grant  NNX08AL48G.  The work of O.K. was supported by RFBR grant 10-02-01406-a. J.F.B. was supported by NSF CAREER Grant PHY-0547102. 

\bibliographystyle{apj}
\bibliography{g}

\end{document}